\documentclass[11pt]{article}

\usepackage{subfigure}
\usepackage[SquareTraceBrackets]{quantum}
\usepackage{graphicx,bm,upgreek,amssymb,amsmath}

\usepackage[T1]{fontenc}
\usepackage[scaled]{helvet}

\DeclareMathOperator{\sinc}{sinc}

\begin{document}

\title{\large\textsf{\textbf{Photonic Quantum Information Processing}}}

\author{
Pieter Kok\\
Department of Physics \& Astronomy, The University of Sheffield}
\date{}

\maketitle

\begin{abstract}
\noindent Information processing with light is ubiquitous, from communication, metrology and imaging to computing. When we consider light as a quantum mechanical object, new ways of information processing become possible. In this review I give an overview how quantum information processing can be implemented with single photons, and what hurdles still need to be overcome to implement the various applications in practice. I will place special emphasis on the quantum mechanical properties of light that make it different from classical light, and how these properties relate to quantum information processing tasks.

\end{abstract}

\section{Information processing with light}
From gestures and smoke signals to books, paintings, telecom and high-capacity optical fibres connecting continents, light is one of the main information carriers that has been driving human civilisation since antiquity. 
Aside from the cultural aspect of communicating ideas with pictures, optical information processing is an important economic engine. In 2020, the market volume of the photonics industry will be 650 billion euro and is expected to continue to outperform GDP \cite{photonics13}.  Light is the ideal medium for fast, reliable and high-bandwidth communication. The amount of data that can be transmitted through optical fibres continues to grow, and we are approaching the limit for the capacity of single-mode fibres. To increase the capacity, multi-mode fibres can be employed to achieve a data transmission rate of 255 Terabits per second over a distance of one kilometre \cite{vanuden14}. For comparison, this is similar to streaming one hundred thousand full length HD movies \emph{each second}. 

Similarly, imaging is fundamentally an optical task, and it has been known for a long time that the wave nature of light places a limit on how well we can make out small details using microscopes and telescopes. The resolution limit for imaging is called the Abbe limit, and depends on the opening angle of the aperture of the imaging device, called the \emph{numerical aperture}. The bigger the numerical aperture, the higher the resolution. This is why telescopes are made in larger and larger sizes. In microscopy, where one can have more control over the object that is to be imaged, several techniques have been developed that can beat the Abbe limit (e.g., STED \cite{hell94}, STORM \cite{rust06}, PALM \cite{betzig06,hess06}). These techniques use prior information about the source and selective activation of photo-emitters to achieve so-called super-resolution.

Finally, conventional computing faces some important barriers, such as heat generation and bandwidth limitations. Classical optics can help heat reduction by using passive elements and reversible computing, and optics also allows us to handle complex calculations by using fan-in and fan-out, in addition to parallelisation \cite{shlomi10}. These techniques are expected to become more prevalent in the near future.

All of the above examples are using classical optics. However, light is not classical. From a fundamental physics point of view, optical information processing must be extended to take quantum effects into account. These effects do not just add noise to existing techniques, but enable dramatic improvements in information processing with light, from communication, metrology and imaging to full-scale quantum computing.

In this review, I will sketch the physical principles and phenomena that lie at the heart of optical quantum information processing. I place special emphasis on the quantum mechanical properties of light that make it different from classical light, and how these properties relate to information processing tasks. In section~\ref{sec:transition} I introduce the concepts of coherence, anti-bunching and the Hong-Ou-Mandel effect, and in section~\ref{sec:photons} I give a brief introduction how photons as quantum information carriers are described mathematically. Section~\ref{sec:comm} is devoted to quantum communication, covering the no-cloning theorem, quantum key distribution, teleportation and repeaters.  In section~\ref{sec:metro} I introduce the idea of precision metrology with classical and quantum light and show how similar techniques can be used for high resolution imaging. Section~\ref{sec:comp} starts with a discussion about optical entanglement and introduces the famous \emph{KLM protocol} for quantum computing. I also sketch the most recent ideas for creating a quantum computer architecture based on linear optics. Finally, in section~\ref{sec:future} I give an overview of the practical challenges that still remain in implementing the ideas presented in this review.

\section{From classical light to quantum light}\label{sec:transition}
As we continue to improve the information processing capabilities of our computer processors and networks, everything is being made smaller and more efficient. In miniaturizing and squeezing all the information out of every last bit of light, we will be coming up against a fundamental limit of nature, namely the fact that light comes in discrete quanta called \emph{photons}. Entirely new laws of physics come into play that have a profound effect on our capabilities for information processing. Before we can explore these new capabilities, however, we need to establish in some more detail what makes quantum light different from classical light. To this end we will briefly describe the idea of coherence, anti-bunching, and two-photon interference.

\subsection{Coherence}\label{sub:coherence}
Classical light exists in roughly two categories, namely \emph{thermal} and \emph{coherent} light. Thermal light is the type that is emitted by sources like stars, light bulbs and LEDs, while coherent light is typically associated with the output of a laser. The two types should not be seen as  completely distinct, but rather as the extremes of a whole spectrum that spans from coherent, via partially coherent, to thermal (or incoherent) light. A key concept in this regard is the \emph{coherence length} of a beam of light. 

\begin{figure}[t!]
\begin{center}
\begin{minipage}{110mm}
\includegraphics[height=33mm]{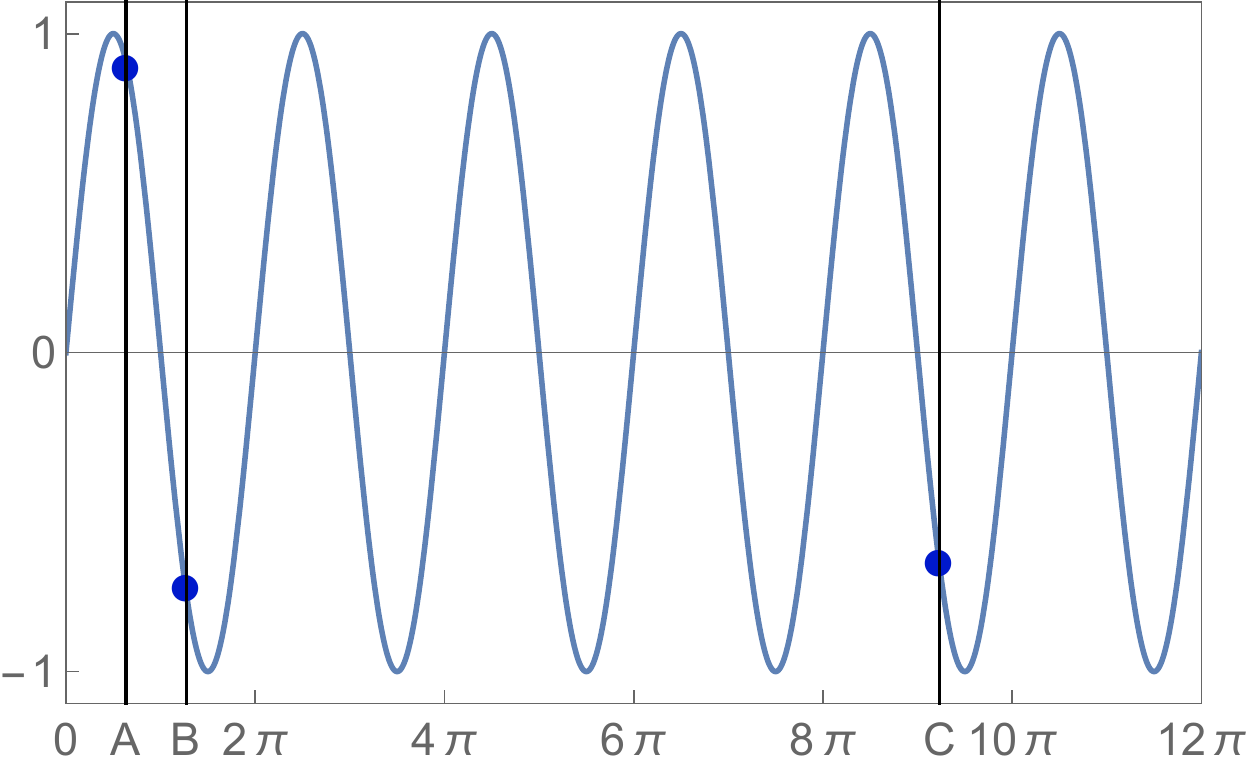} \hfill
\includegraphics[height=33mm]{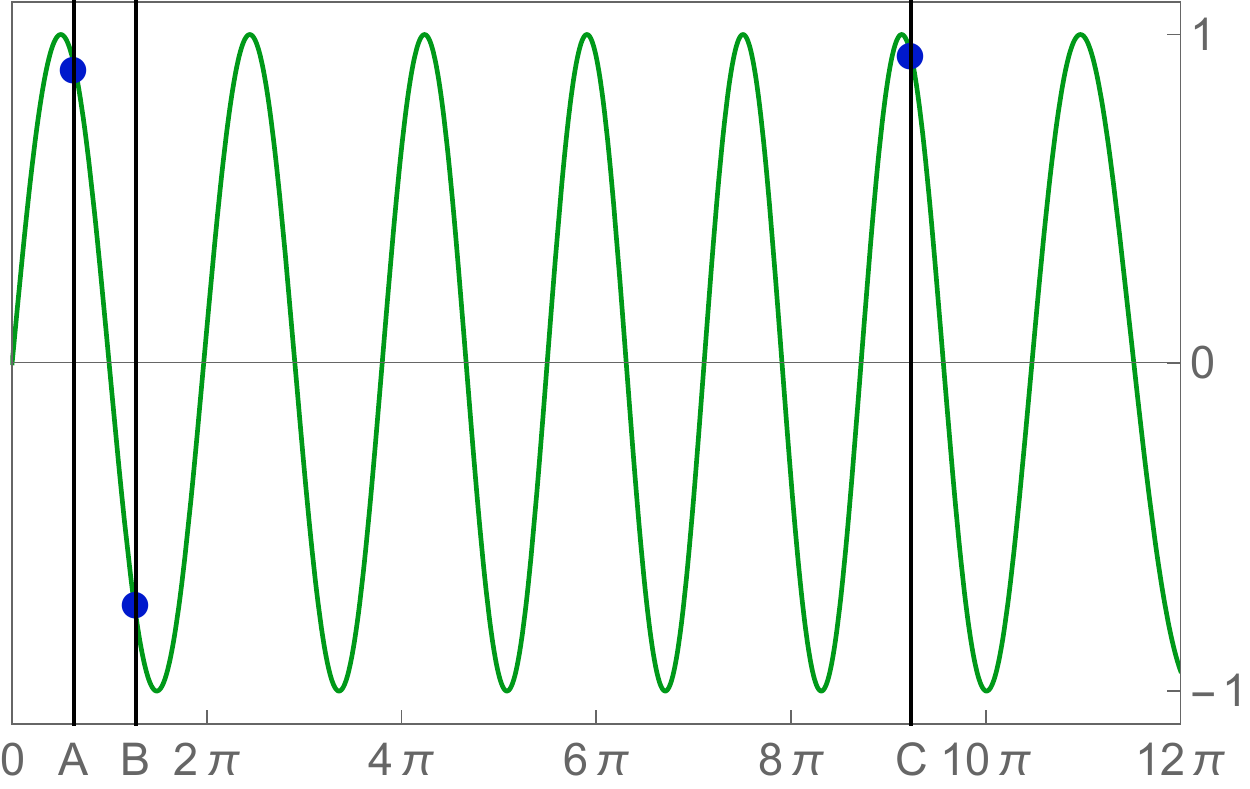}
\caption{Coherence: A wave of constant frequency over a long time (left figure) will have a well-defined phase relationship both at short times (between point $A$ and point $B$), as well as long times (between point $A$ and point $C$). Alternatively, if the frequency fluctuates over time (right figure), there will still be an appreciable phase correlation at short times (between point $A$ and point $B$), but at long times the phase relationship fluctuates (between point $A$ and point $C$). The characteristic time at which the coherence between the phases drops below a certain level is called the coherence time. Sometimes this is also referred to as the coherence length.} \label{fig:coherence}
\end{minipage}
\end{center}
\end{figure}

The coherence length is most easily explained by first looking at the coherence \emph{time}. Consider a wave of constant frequency $\omega$, as shown in figure \ref{fig:coherence} on the left. The phase of the wave at time $t$ will be given by $\phi$. We can calculate the difference between the phases at difference times. For example, $\phi_B - \phi_A$ is the difference in the phase of the wave for the time interval $\Delta t = t_B - t_A$. Since the frequency of the wave is constant over time, the phase difference $\phi_B - \phi_A$ between two points in time separated by $\Delta t$ will be constant as well. In addition, given a third phase $\phi_C$ at time $t_C$ much later than $t_A$ we have again a constant phase difference $ \phi_C - \phi_A$ for a time interval $\Delta T = t_C - t_A$:
\begin{align}
 \phi_B - \phi_A = \text{constant} \qquad\text{and}\qquad \phi_C - \phi_A= \text{constant.} 
\end{align}
Next, consider a wave whose frequency fluctuates over time, as shown in figure \ref{fig:coherence} on the right (the effect is quite subtle). If the time interval $\Delta t = t_B - t_A$ is short compared to the (inverse) rate of change of $\omega$, the difference between the phases at $A$ and $B$ will still be nearly constant. However, for longer times between two points on the wave this will no longer be the case. The fluctuations in the frequency will cause fluctuations in the phase relationship between $A$ and $C$, and we have
\begin{align}
 \phi_B - \phi_A \simeq \text{constant} \qquad\text{and}\qquad \phi_C - \phi_A = f(t)\, ,
\end{align}
where $f(t)$ is a strongly fluctuating function over time taking values in the entire interval $[0,2\pi)$. There will be a characteristic time $\tau$ where the behaviour of the phase difference changes from nearly constant to strongly fluctuating. This is the coherence time of the wave. Multiplying the coherence time by the velocity of the wave in the medium gives us the coherence length. A very similar argument can be made for the phase coherence of two emitters a distance $d$ apart, leading to the concept of \emph{transverse} coherence length.

The coherence length of light plays a fundamental role in classical interference. Coherent sources  interfere, while incoherent sources do not. These properties carry over to the quantum mechanical treatment of light, and we will see that coherence is a crucial property for optical quantum information tasks.

\subsection{Anti-bunching}
Classical waves can have any amplitude, no matter how small. However, this is not the case for quantum mechanical light. When the power of a light source is reduced, at some point a detector will no longer record a constant signal, but rather we find that the light arrives in bursts. These bursts are called photons. 

We can easily imagine how such bursts come about: the atom that emits the photon does so by having one of its electrons drop down to a lower energy state (the difference in energy escapes in the form of our photon). Creating a second burst of light requires the electron to be loaded back up into the excited state, which takes time. This leads to the phenomenon of \emph{anti-bunching}: it is relatively unlikely that two photons are detected in much shorter succession than the time it takes the electron to reoccupy the excited state.

To make this description a bit more mathematically precise, we can consider the probability distribution of the number of photons that are recorded in a time interval $\Delta t$. If the photons arrive completely randomly, this distribution will be Poissonian: 
\begin{align}\label{eq:poisson}
 \pr{n} = \frac{\lambda^n {\rm e}^{-\lambda}}{n!}\, ,
\end{align}
where $n$ is the number of photons detected in the interval $\Delta t$ and $\lambda$ is the (dimensionless) intensity of the light, which corresponds to the average number of photons in $\Delta t$. When a source exhibits anti-bunching at a time scale $\Delta t$, the probabilities of finding two, three, four, etc., photons will be suppressed compared to the Poisson distribution in Eq~(\ref{eq:poisson}).

\begin{figure}[t!]
\begin{center}
\begin{minipage}{110mm}
\includegraphics[height=41mm]{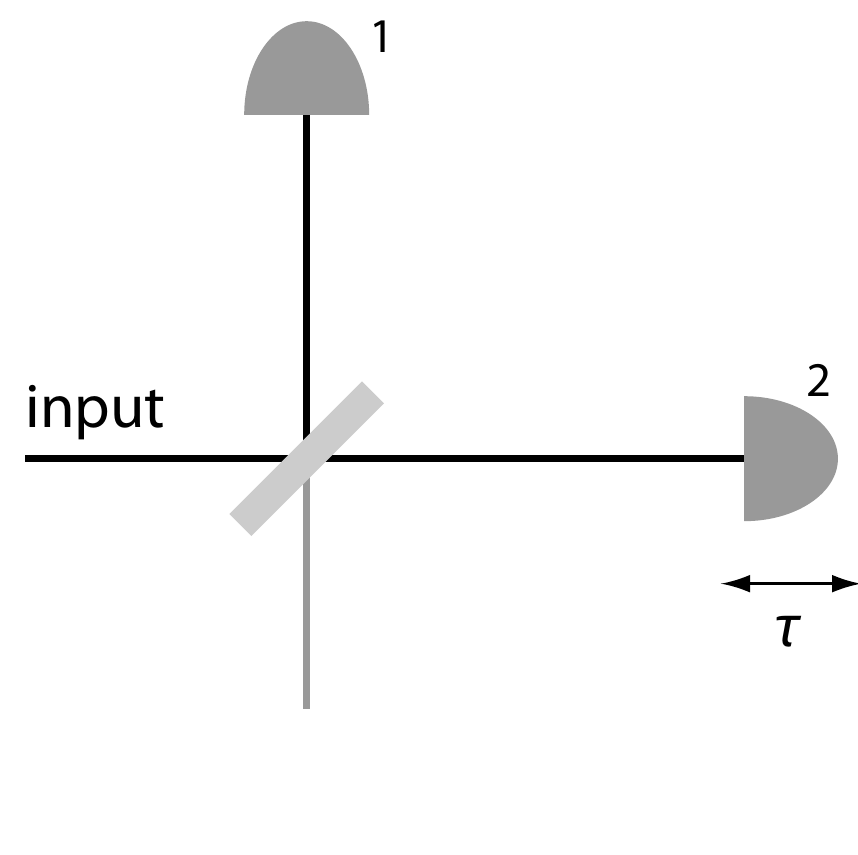} \hfill
\includegraphics[height=41mm]{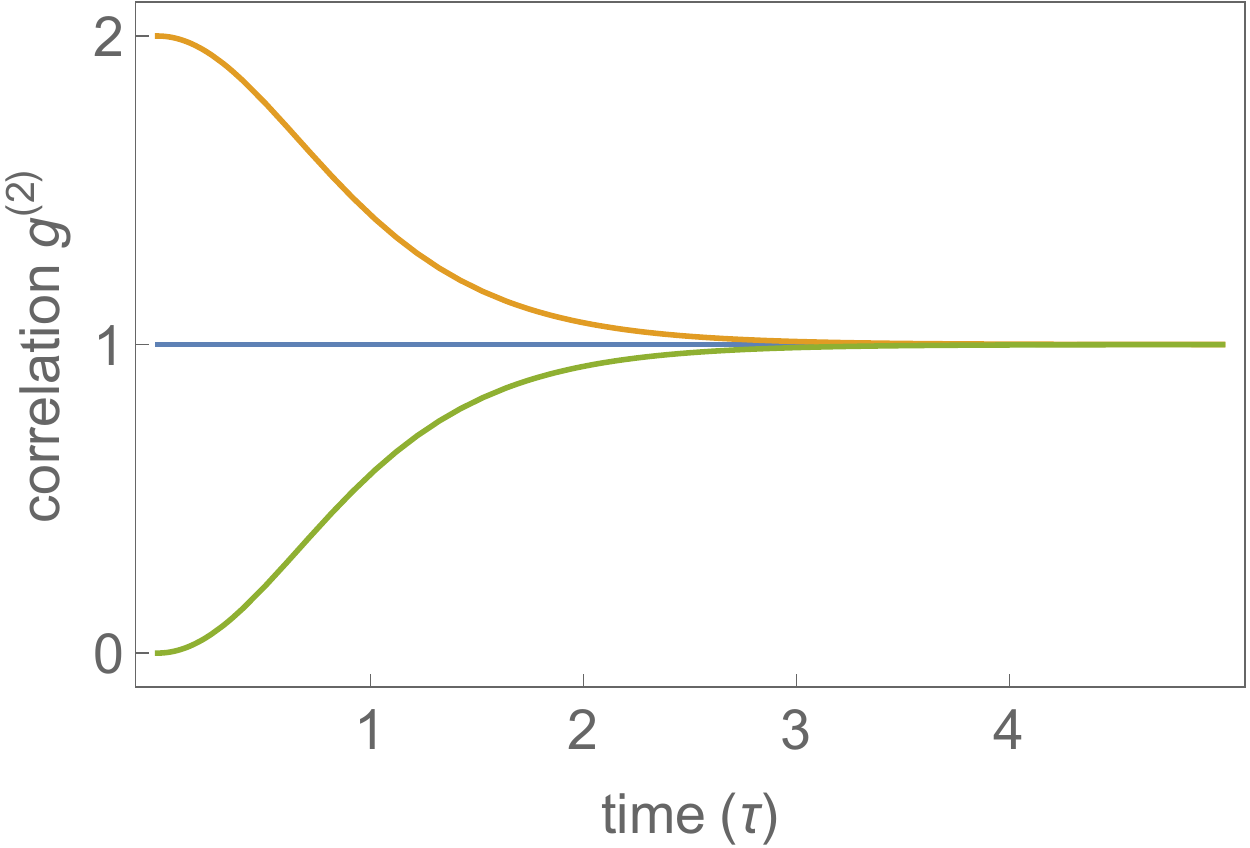}
\caption{Second-order correlations: By counting the number of coincidence detections in detectors 1 and 2 we can determine the second order correlation function $g^{(2)}$. We vary the path length of one of the detectors. For a Poissonian distribution (blue curve) the $g^{(2)}$ function remains constant at 1, while at very short times (i.e., equal distances of the detectors to the beam splitter) the probability of getting detector coincidences drops to zero for anti-bunched light (green curve). Note that thermal light is bunched (orange curve), in the sense that it is \emph{more} likely that two photons are emitted at the same time.} \label{fig:hbt}
\end{minipage}
\end{center}
\end{figure}

Experimentally, we can demonstrate this by putting a beam splitter in the light beam and count the number of coincidences in the two photodetectors, as shown in figure \ref{fig:hbt}. The number operator for detector 1 is given by $\hat{n}_1$ and that of detector 2 by $\hat{n}_2$. The average photon number in detector $j=1,2$ at time $t$ is then calculated via the quantum mechanical expectation value $\braket{\hat{n}_j(t)}$. The second-order correlation is measured by the $g^{(2)}$ function defined according to 
\begin{align}
 g^{(2)}(\tau) = \frac{\Braket{\hat{n}_1(t)\;\hat{n}_2(t+\tau)}}{\Braket{\hat{n}_1(t)}\Braket{\hat{n}_2(t+\tau)}}\, ,
\end{align}
which does not depend on $t$ for stationary processes, such as the situation we consider here. A typical $g^{(2)}(\tau)$ function is shown in figure \ref{fig:hbt}. For long time scales the intensities for anti-bunched light in both detectors become independent, and the $g^{(2)}$ function for anti-bunched light (the green curve) tends towards that of Poissonian light (the blue curve). For completeness we also plotted the $g^{(2)}$ function for thermal light (the orange curve), which shows \emph{bunching}: it is more likely that the two detectors fire in unison at very short timescales compared to Poissonian light. This is in fact also a quantum mechanical effect: a thermal ``gas'' of photons obeys the Bose-Einstein distribution, rather than the classical Maxwell-Boltzmann distribution.

\subsection{The equations of motion for a photon}
The question is now what makes light quintessentially quantum-mechanical, and which features of light are well-explained by the classical theory. Maxwell's equations do not provide a description of quantised energy, but they do very accurately describe the shapes of the wave packets. In particular, the coherence lengths (transverse and longitudinal) that are determined by the classical theory determine the interference properties of photons. To illustrate this, we consider Young's double-slit experiment with photons.

Suppose that a source of single photons illuminates a screen with two narrow slits of width $a$ placed at a distance $d$ apart. The classical theory predicts that the intensity pattern at a screen a distance $L$ from the slits is given by 
\begin{align}
 I(x) = I_0 \cos\left(\frac{xd\pi}{L\lambda}\right) \sinc\left(\frac{xa\pi}{L\lambda}\right) \qquad\text{with}\qquad \sinc\beta \equiv \frac{\sin\beta}{\beta}\, ,
\end{align}
where $x$ is the position along the screen and $\lambda$ is the wavelength of the light. When the source emits single photons one at a time, after collecting many photons the average intensity on the screen will be exactly the same. Therefore the spatial behaviour of single photons is identical to that of classical waves. However, this presupposes that we do not know which slit the photon travels through. This is equivalent to saying that the amplitudes of the photon wave packet at the left slit and the right slit add coherently. In other words, the transverse coherence of the photon wave packet must be larger than the distance between the slits.

Next, we consider the quantum mechanical description of light. Classically, we can write the electric field $E$ as a solution to the wave equation:
\begin{align}\label{eq:classicalfieldop}
 E(\mathbf{x},t) = \sum_{\mathbf{k}} A_ {\mathbf{k}}\, u_ {\mathbf{k}} (\mathbf{x},t) + A_ {\mathbf{k}}^*\, u_ {\mathbf{k}}^* (\mathbf{x},t)\,  ,
\end{align}
where $A$ is the complex amplitude of a wave in mode $\mathbf{k}$ (commonly referred to as the wave vector, but in principle we can have more exotic labelings), and the sum over $\mathbf{k}$ indicates that the waves can be superposed. The functions $u_ {\mathbf{k}} (\mathbf{x},t)$ in Eq.~(\ref{eq:fieldop}) are the so-called \emph{mode functions}, and they form a complete set of solutions to the classical Maxwell equations. Often a plane wave expansion for ${E}(\mathbf{x},t)$ is given, in which  
\begin{align}
% u_ {\mathbf{k}} (\mathbf{x},t) = \sqrt{\frac{\hbar  \omega_{\mathbf{k}}}{2\epsilon_0 (2\pi)^3}}\, \bm{\epsilon}_\mathbf{k}\, {\rm e}^{i\mathbf{k}\cdot\mathbf{x} - i\omega_{\mathbf{k}}t} ,
 u_ {\mathbf{k}} (\mathbf{x},t) \propto \bm{\epsilon}_\mathbf{k}\, {\rm e}^{i\mathbf{k}\cdot\mathbf{x} - i\omega_{\mathbf{k}}t} ,
\end{align}
and the sum in Eq.~(\ref{eq:fieldop}) becomes an integral over $\mathbf{k}$ with $\omega_\mathbf{k}$ the frequency of the wave vector $\mathbf{k}$. The vector $ \bm{\epsilon}_\mathbf{k}$ determines the polarisation of the mode. Other expansions are also possible, and may be more convenient depending on the application (for example, a plane wave expansion is not very suited to describe a wave in an optical fibre). 

In the full quantum mechanical description of optics, the electric field becomes an operator and can be written as \cite{koklovett10}:
\begin{align}\label{eq:fieldop}
 \hat{E}(\mathbf{x},t) = \sum_{\mathbf{k}} \hat{a}_ {\mathbf{k}}\, u_ {\mathbf{k}} (\mathbf{x},t) + \hat{a}_ {\mathbf{k}}^\dagger\, u_ {\mathbf{k}}^* (\mathbf{x},t)\,  ,
\end{align}
where $\hat{a}_ {\mathbf{k}}$ and $\hat{a}_ {\mathbf{k}}^\dagger$ are the annihilation and creation operator for the optical mode indicated by $\mathbf{k}$. These operators replace the complex amplitudes in the classical theory and obey the commutation relations
\begin{align}
 \left[ \hat{a}_ {\mathbf{k}}, \hat{a}_ {\mathbf{k}'} \right] = \left[ \hat{a}_ {\mathbf{k}}^\dagger, \hat{a}_ {\mathbf{k}'}^\dagger \right] = 0 \qquad\text{and}\qquad \left[ \hat{a}_ {\mathbf{k}}, \hat{a}_ {\mathbf{k}'}^\dagger \right] = \delta_{\mathbf{k}\, \mathbf{k}'}\, ,
\end{align}
where $\delta_{\mathbf{k}\, \mathbf{k}'}$ is the Kronecker delta symbol\footnote{We can also define the electric field operator over a continuum $\mathbf{k}$, in which case the sum over $\mathbf{k}$ becomes an integral and the Kronecker delta becomes a Dirac delta $\delta(\mathbf{k} - \mathbf{k}')$.}. The creation and annihilation operators act on photon number states $\ket{n}_\mathbf{k}$ according to the rules
\begin{align}
 \hat{a}_{\mathbf{k}} \ket{n}_{\mathbf{k}} = \sqrt{n} \ket{n-1}_{\mathbf{k}} \qquad\text{and}\qquad \hat{a}_{\mathbf{k}}^\dagger \ket{n}_{\mathbf{k}} = \sqrt{n+1} \ket{n+1}_{\mathbf{k}}\, .
\end{align}
The number operator for mode $\mathbf{k}$ is then given by $\hat{n}_{\mathbf{k}} = \hat{a}_{\mathbf{k}}^\dagger \hat{a}_{\mathbf{k}}$. The algebra of these operators is identical to that of the simple harmonic oscillator.

Creating a photon in mode $\mathbf{k}$ means that the photon will behave according to the spatio-temporal description provided by $u_ {\mathbf{k}} (\mathbf{x},t)$. Since $u_ {\mathbf{k}} (\mathbf{x},t)$ is determined by Maxwell's equations, we can say that the classical Maxwell equations are the equations of motion for the photon. The quantum mechanical behaviour of light is restricted to the photon \emph{statistics}.

\subsection{The Hong-Ou-Mandel effect}
\begin{figure}[t!]
\begin{center}
\begin{minipage}{110mm}
\includegraphics[width=110mm]{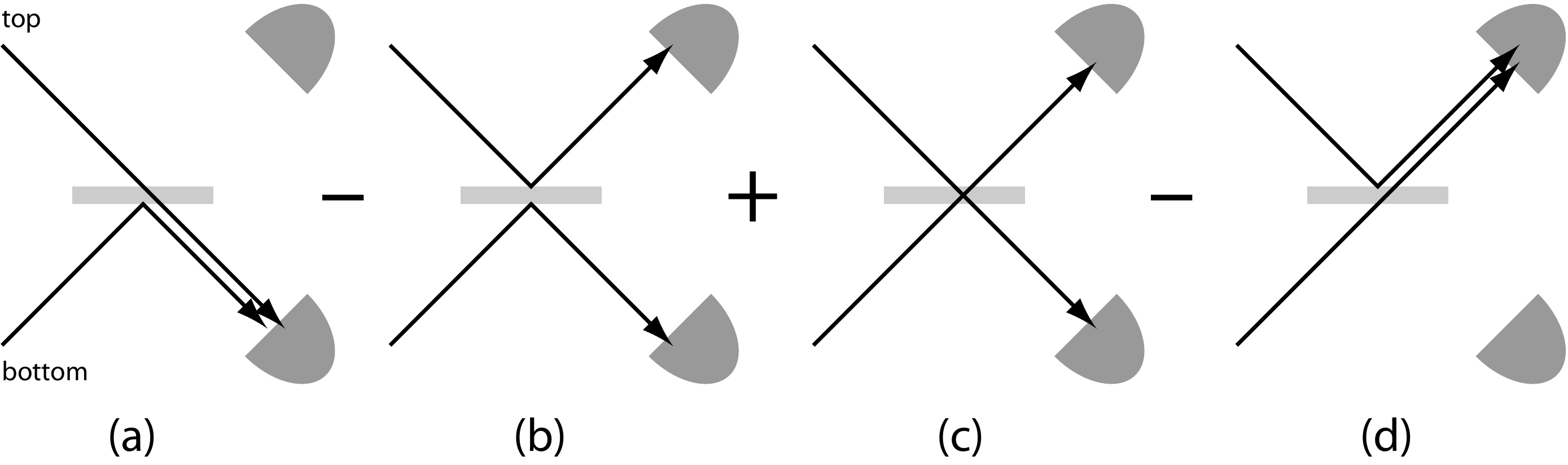}
\caption{The Hong-Ou-Mandel effect: If we send two photons into the two input beams of a 50:50 beam splitter there are four possible outcomes, as shown in the figure. In quantum mechanics, these four possibilities are superposed. Due to a $\pi$ phase shift upon reflection of the top surface, the case where both photons are reflected has a relative minus sign compared to the case where both photons are transmitted. When the input photons are identical, we cannot distinguish between the two middle outcomes, and they cancel. So the identical photons will always pair off towards the same output beam, and never leave the beam splitter in different beams.} \label{fig:hom}
\end{minipage}
\end{center}
\end{figure}

While the mode shapes of propagating photons are determined by the classical theory of electrodynamics, the quantum behaviour of light is most apparent in multi-photon effects. For the purposes of quantum information processing, the most important example of this is the Hong-Ou-Mandel effect, which is a two-photon intensity interference effect. Specifically, the Hong-Ou-Mandel effect occurs when two photons with identical frequency, polarisation and shape of the wave packet enter a 50:50 beam splitter on either side. If we place detectors in the two outgoing modes of the beam splitter, each photon has two ways of triggering the detectors. Either the photon triggers the upper detector, or it triggers the lower detector. The resulting four possible paths for the two photons are shown in figure~\ref{fig:hom}. In (a) the top input photon is transmitted while the bottom photon is reflected. In that case, both photons end up in the bottom detector. And vice versa, both photons may end up in the top detector (d). Whenever a photon is reflected off the topside of the beam splitter it experiences a $\pi$ phase shift, which results in a factor $e^{i\pi} = -1$ in the state of the photon.

\begin{figure}[b!]
\begin{center}
\begin{minipage}{110mm}
\includegraphics[width=80mm]{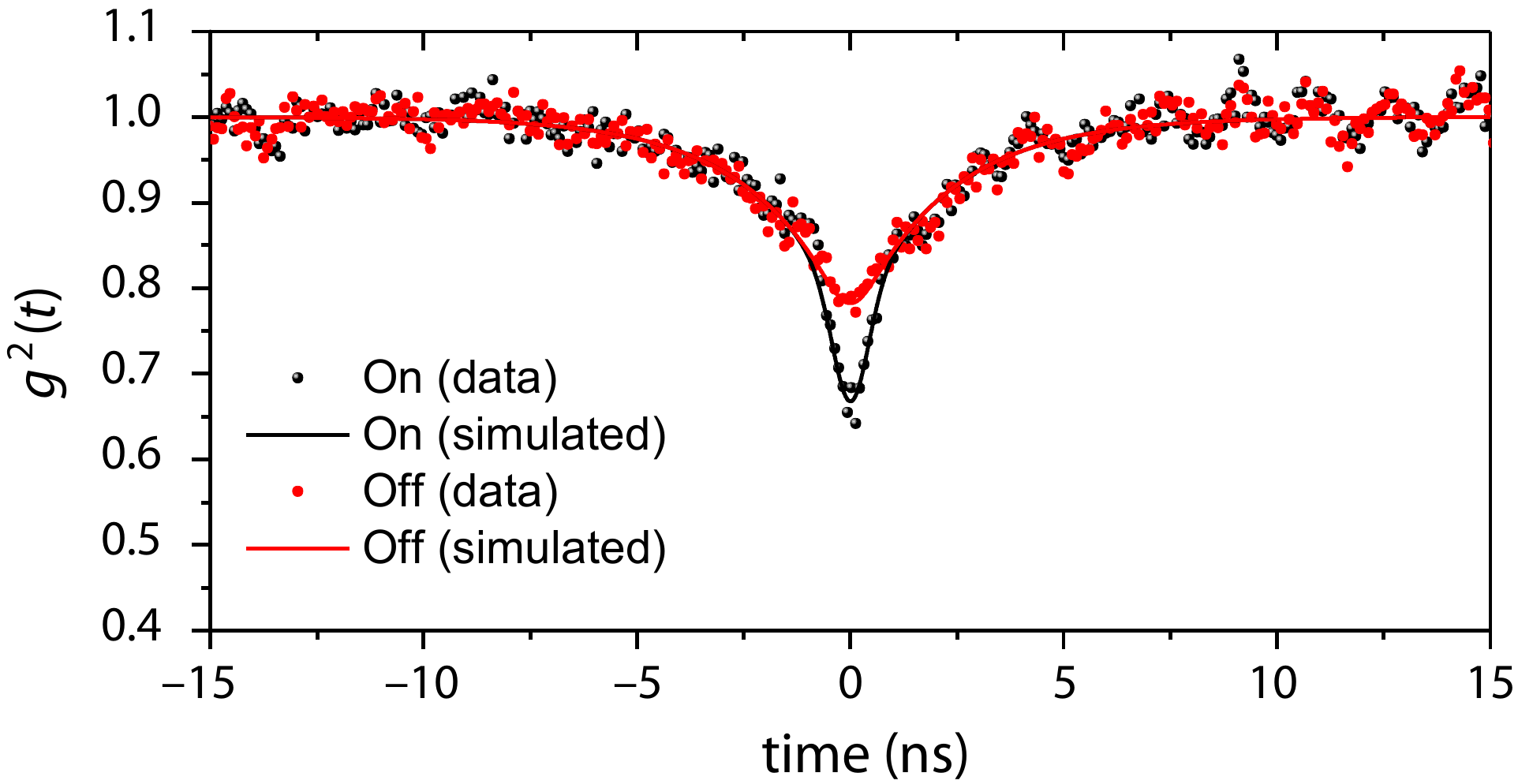}
\caption{Experimental data for the Hong-Ou-Mandel effect using one dot and one laser input \cite{prtljago16}. The dip is not as deep as one would expect for two indistinguishable single photon sources, because the laser is not a single-photon source (it obeys Poissonian statistics, rather than anti-bunching). Nevertheless, the dip (black curve) is lower than expected for classical light (red curve).} \label{fig:homexp}
\end{minipage}
\end{center}
\end{figure}

An interesting effect happens when we consider situations (b) and (c) in figure \ref{fig:hom}. In case (b), both photons are reflected by the beam splitter, while in case (c), both photons are transmitted. Since the photons are identical, the two processes (b) and (c) are \emph{indistinguishable} from each other. No physical process can tell whether the photons were reflected or transmitted, and the beam splitter itself holds no memory of the process. According to Feynman, this means that the two contributions must be superposed coherently and are allowed to interfere \cite{feynman}. However, since the top photon in process (b) picks up a phase factor $e^{i\pi}$, the two states must be subtracted. This leads to destructive interference, and as a result the two identical photons will \emph{never} end up in separate detectors. This effect was first observed by Hong, Ou, and Mandel in 1987 \cite{hong87}. The effect lies at the heart of the protocols that enable quantum computing with single photons and linear optics. A recent experimental realisation of the Hong-Ou-Mandel effect is shown in figure \ref{fig:homexp} \cite{prtljago16}. The depth of the dip indicates the level of indistinguishability between the two photons. Together with a single-mode $g^{(2)}$ measurement to verify the presence of only a single photon, the Hong-Ou-Mandel dip gives a good indication for the quality of single-photon sources.

\section{Photons as quantum information carriers}\label{sec:photons}
As we encounter the quantum limits of light, we may ask what we can do in terms of information processing if we embrace this natural behaviour. Imagine that we send classical bits using the polarisation degree of freedom. In other words, an optical pulse of horizontally polarised light ($H$) is defined as the bit value 0, and a vertically polarised pulse ($V$) is the bit value 1. At the single photon level the polarisation is still a well-defined physical property, since it is determined by the mode functions (and therefore obey Maxwell's equations). The polarised photons and their bit values can be described quantum mechanically with the quantum states 
\begin{align}
 \ket{H} \equiv \ket{0} \qquad\text{and}\qquad \ket{V} \equiv \ket{1}\, .
\end{align}
A fundamental property of classical light is that two pulses can be prepared in superposition. For our example, this means that we can make a superposition of vertical and horizontal light. The result is a new pulse with a polarisation in a different direction depending on the relative phase between the $H$ and $V$ pulses.  The new polarisation can be linear, circular, or elliptical. 

This property carries over to photons. For example, left- and right-handed circularly polarised photons have quantum states $\ket{L}$ and $\ket{R}$, respectively:
\begin{align}
 \ket{L} = \frac{\ket{H}+i\ket{V}}{\sqrt{2}} \qquad\text{and}\qquad \ket{R} = \frac{\ket{H}-i\ket{V}}{\sqrt{2}} \, .
\end{align}
If we treat the horizontal and vertical polarisation of a photon as bit values 0 and 1, we see that we now have two \emph{new} bit values that are superpositions of $\ket{0}$ and $\ket{1}$:
\begin{align}
 \ket{\circlearrowleft} = \frac{\ket{0}+i\ket{1}}{\sqrt{2}} \qquad\text{and}\qquad \ket{\circlearrowright} = \frac{\ket{0}-i\ket{1}}{\sqrt{2}} \, .
\end{align}
This is not possible for a classical bit, and we therefore call the polarised photon a quantum bit, or \emph{qubit} for short. Every classical polarisation has a corresponding qubit when we bring the optical pulse down to a single photon. This requires that the two pulses have a well-defined phase relationship, and are therefore \emph{coherent} in the sense of the discussion in section~\ref{sub:coherence}. The extra states in a qubit over a classical bit suggest that information processing with qubits can be more powerful than information processing with classical bits, because---loosely speaking---more available states means more room for the information to play in. Instead of polarisation, we can use other degrees of freedom of light \cite{koklovett10}, but for simplicity we will restrict our discussion to polarised photons in this article.

Since the qubit structure is directly inherited from the classical superposition principle, the next question is: what makes the photonic qubit a fundamentally quantum mechanical object? The answer is given by anti-bunching. There is only one indivisible photon (see figure \ref{fig:hbt}) that triggers either detector 1 or detector 2. Classically, both detectors could register a non-zero signal simultaneously. The fact that this is not possible for single photons means that the photon ends up in either one or the other detector in a \emph{probabilistic} manner. If we wish to measure the polarisation of the photon, we must first choose which \emph{basis} we want to measure ($\ket{H}$ and $\ket{V}$, or $\ket{L}$ and $\ket{R}$). If we measure the photon in the state $\ket{L}$ in the $H/V$ basis, we will obtain the outcome ``H'' or ``V'' with 50:50 probability. 

The superposition principle---together with the concept of the photon as a particle---further leads to the phenomenon of \emph{entanglement}. Two photons can be prepared in the state 
\begin{align}\label{eq:phiplus}
 \ket{\Phi^+} = \frac{\ket{HH} + \ket{VV}}{\sqrt{2}}\, ,
\end{align}
where $\ket{HH}\equiv\ket{H}_1\ket{H}_2$ and $\ket{VV}\equiv\ket{V}_1\ket{V}_2$ are short-hand for the polarisation states of the two photons.
It is easy to see that the state in Eq.~(\ref{eq:phiplus}) cannot be written as the product of two separate photon states:
\begin{align}
 \ket{\Phi^+} \neq \left( \alpha_H \ket{H} + \alpha_V \ket{V} \right)  \left( \beta_H \ket{H} + \beta_V \ket{V} \right)\, ,
\end{align}
where $\alpha$ and $\beta$ are complex numbers obeying $\abs{\alpha_H}^2 + \abs{\alpha_V}^2 = 1$ and $\abs{\beta_H}^2 + \abs{\beta_V}^2 = 1$. The effect of entanglement is that the two photons are more strongly correlated than is possible classically:
\begin{align}
 \ket{\Phi^+} & = \frac{\ket{HH} + \ket{VV}}{\sqrt{2}} = \frac{1}{2\sqrt{2}} \left[ (\ket{L}+\ket{R}) (\ket{L}+\ket{R}) + (-i)^2 (\ket{L}-\ket{R}) (\ket{L}-\ket{R}) \right] \cr & = \frac{\ket{LR} + \ket{RL}}{\sqrt{2}}\, .
\end{align}
There is not only a correlation in $H$ and $V$, but also in $L$ and $R$. This is not possible in classical systems, and these stronger quantum correlations can be utilised in information processing. For future use, we define a basis of four entangled states, called the \emph{Bell states}:
\begin{align}
 \ket{\Phi^+} = \frac{\ket{HH} + \ket{VV}}{\sqrt{2}} & \qquad\phantom{\text{and}}\qquad \ket{\Psi^+} = \frac{\ket{HV} + \ket{VH}}{\sqrt{2}} \, ,\cr 
 \ket{\Phi^-} = \frac{\ket{HH} - \ket{VV}}{\sqrt{2}} & \qquad\text{and}\qquad \ket{\Psi^-} = \frac{\ket{HV} - \ket{VH}}{\sqrt{2}} \, .
\end{align}
A measurement in this basis is called a \emph{Bell measurement}, and plays a crucial role in quantum information processing.

\bigskip

\noindent
Now that we have a photonic qubit, what exactly can we do with it? We would expect that all the classical information processing tasks with light will in some way carry over to quantum light with some enhancements due to qubit superpositions and entanglement. Indeed, we can construct new communication protocols and create a quantum internet \cite{kimble08}, we can use photons as measurement probes to achieve a much higher precision in parameter estimation \cite{giovannetti11} and imaging \cite{boto00,kolobov07}, and we can use photons as the fundamental information carriers in quantum computing \cite{kok07}.

However, photons are not equally good at all these things. While they are clearly very good data carriers over long distances, it is rather hard to slow them down significantly or even stop them completely. Typical classical and quantum information processing tasks require feed-forward operations in which the state of a qubit is modified in a way that depends on the measurement outcome of another process. If the photon flies away at the speed of light while the measurement is being made, we cannot perform feed-forward because we cannot catch up with the photon. We therefore need to store the photon in some kind of photon memory. This is a complicated process that will likely introduce a lot of noise.

Another complication is that while the polarisation of a single photon is very easy to manipulate using half wave plates and quarter wave plates, the creation of entanglement between two photons is extremely hard. This is due to the complete absence of direct photon-photon interactions. An operation that entangles two photons must therefore be an inherently nonlinear process, either involving nonlinear materials or a clever arrangement of projective measurements. In this review we will consider the latter.

\section{Quantum communication}\label{sec:comm}
Classical light makes for an excellent information carrier over long distances. This is also true for quantum light. Moreover, we can use the quantum mechanical properties of photons to accomplish new communication tasks that are more difficult or impossible with classical light. As an example, we will consider secure communication using quantum key distribution, and explore what requirements are necessary to extend these techniques beyond a few hundred kilometres.

\subsection{The no-cloning theorem and quantum key distribution}
Consider a photon with horizontal and vertical polarisation states $\ket{H}$ and $\ket{V}$, respectively. As we have seen, we can make quantum superpositions of these states to obtain different polarisation states. The no-cloning theorem says that it is impossible to create a machine that copies an unknown quantum state perfectly \cite{wootters82,dieks82}. To see this, suppose that we have a photon in some unknown polarisation state $\ket{\psi}$ and a second photon in a known initial polarisation state, e.g., $\ket{H}$. A proper copying machine would have to produce the following effect on \emph{any} state $\ket{\psi}$:
\begin{align}
 \ket{\psi}\ket{H} \underset{\rm copy}{\longrightarrow} \ket{\psi}\ket{\psi}\, .
\end{align}
In particular, the machine must act on the states $\ket{H}$ and $\ket{V}$ according to 
\begin{align}
 \ket{H}\ket{H} \underset{\rm copy}{\longrightarrow} \ket{H}\ket{H} \qquad\text{and}\qquad \ket{V}\ket{H} \underset{\rm copy}{\longrightarrow} \ket{V}\ket{V} \, .
\end{align}
This completely determines how the copying machine will handle the unknown polarisation state, because any polarisation state can be written as a quantum superposition of $\ket{H}$ and $\ket{V}$. For example, suppose that the state $\ket{\psi}$ is in fact the left-handed circularly polarised state $\ket{L}$. Then
\begin{align}
 \ket{\psi} = \frac{\ket{H}+i \ket{V} }{\sqrt{2}} \, ,
\end{align}
and the copying machine will produce
\begin{align}
 \ket{\psi}\ket{H} = \frac{\ket{H}+i\ket{V}}{\sqrt{2}} \ket{H}  \underset{\rm copy}{\longrightarrow} \frac{1}{\sqrt{2}}\ket{H}\ket{H}+\frac{i}{\sqrt{2}}\ket{V}\ket{V} \, .
\end{align}
However, this is \emph{not} the same as $\ket{\psi}\ket{\psi}$, as you can tell when we write it out in the $H/V$ basis:
\begin{align}
 \ket{\psi}\ket{\psi} & = \left( \frac{\ket{H}+i\ket{V}}{\sqrt{2}}\right)  \left( \frac{\ket{H}+i\ket{V}}{\sqrt{2}} \right) \cr & = \frac12 \ket{H}\ket{H} + \frac{i}{2}\ket{H}\ket{V} + \frac{i}{2}\ket{V}\ket{H} - \frac12 \ket{V}\ket{V} \, .
\end{align}
This means that a copying machine that works for $\ket{H}$ and $\ket{V}$ will not faithfully copy $\ket{L}$ and $\ket{R}$, and vice versa. Practically, this means that the badly copied photon behaves differently from a photon in the original state. The no-cloning theorem is a fundamental result in quantum mechanics and applies to all physical systems. 

Next, consider the situation in which two agents, Alice and Bob, wish to communicate in private. One way they can accomplish this if they share a secret string of random zeros and ones called a \emph{key}: Alice adds this string to her binary message, creating the encrypted message. Bob decrypts the message by subtracting the key from the encrypted message. Since no-one else has the secret key, nobody can decrypt the message but Alice and Bob. The question is how to generate such a secret key.

Sending the secret key over a public channel will invite eavesdroppers to copy it and gain access to the private message between Alice and Bob. If Alice and Bob can detect the eavesdropper, they know the channel is compromised and move to a different channel. This is what the no-cloning theorem allows them to do: Let's suppose that $\ket{H}$ and $\ket{L}$ denote a bit value of zero, and $\ket{V}$ and $\ket{R}$ denote a bit value of one. Alice sends a random string of photons in polarisation states $\ket{H}$, $\ket{V}$, $\ket{L}$, and $\ket{R}$. Bob measures randomly in the $H/V$ basis or the $L/R$ basis. About half the time Alice will have created a photon in the same basis as Bob's measurement, and in those cases both Alice and Bob will know whether they shared a zero or a one. In the rest of the cases there is no correlation between the bit value sent by Alice and the bit value measured by Bob. Alice and Bob then publicly compare their preparation and measurement bases ($H/V$ or $L/R$), and keep only those bits for which the preparation and measurement bases coincide. Note that they do not reveal the actual bit values, only the bases.

To see whether there is an eavesdropper on the line, Alice and Bob sacrifice a small part of their secret key. They publicly compare this part of the key and see if the bit values match up. If they do, there was no eavesdropper, but if there is a sizeable amount of errors there may have been an eavesdropper. Anyone trying to copy the secret key as it was being established must copy the information of the photon polarisation. However, since the photons were sent in two different bases (unknown to anyone but Alice), a copying machine that works perfectly in the $H/V$ basis will create imperfect copies and cause incorrect measurement results for Bob. These incorrect measurement results will show up when Alice and Bob compare part of their secret key. The secrecy of the key is therefore guaranteed by the no-cloning theorem.

The comparison between Alice and Bob of a fraction of their secret key is what guarantees the privacy of the key. There is a trade-off between the amount of information Eve can gain, and the level of privacy attained by Alice and Bob. When this protocol is implemented with real devices, additional noise will appear in the system, and Alice and Bob must be able to account for that also. To this end, they can further sacrifice part of the key to increase their privacy. This is called \emph{privacy amplification} \cite{koashi07}, and is a crucial part of any practical implementation of quantum key distribution. The general trade-off in the communication between Alice and Bob is then privacy versus bit rate.

\bigskip

\noindent
A final observation about this quantum key distribution protocol is that it relies critically on the quantum mechanical possibility of anti-bunching of light. If light did not come in discrete packages (photons), the eavesdropper could siphon off a small part of the signal with a beam splitter and measure the polarisation of this very weak field. The fact that there is only one photon in each successive pulse from Alice means that it either shows up in Bob's detector, or it arrives in the eavesdropper's detector, in which case Bob will register a failed transmission. In either case that photon will not be used for the secret key and is useless to the eavesdropper. If there was a possibility of more than one photon in each pulse (a Poissonian or thermal distribution), the eavesdropper could measure one photon while an identical photon makes its way to Bob. Note that this does not violate the no-cloning theorem, since Alice can create however many photons she chooses in a state of her choice. 

\subsection{Quantum repeaters and quantum memories}
When a photon travels in an optical fibre, it has a certain probability of being scattered by impurities in the fibre. In this case the photon does not make it to the end of the fibre, and we call this \emph{photon loss}. A fibre can be characterised by an attenuation length $\ell$ at which the original signal is reduced by a factor $1/e$. The attenuation for a fibre of length $L$ is then given by $\exp(-L/\ell)$. This is an exponential decay, which means that we cannot lay arbitrarily long fibre-optic cables and still expect a sizeable bit rate from end to end. In practice, the cable length can be a few hundred kilometres at most.

If we want to extend the reach of quantum communication protocols, we have to add some active devices in the communications channel. Classically, this is accomplished by repeater stations, which amplify the signal and transmit it to the next repeater station. However, amplification is a form of copying, and we have just seen that the no-cloning theorem prevents such devices from working properly on general qubit states. At first this looks like quantum communication will remain viable only for short distances. However, another fundamental protocol in quantum mechanics comes to the rescue here, namely quantum \emph{teleportation} \cite{bennet93}.

\begin{figure}[t!]
\begin{center}
\begin{minipage}{110mm}
\includegraphics[height=40mm]{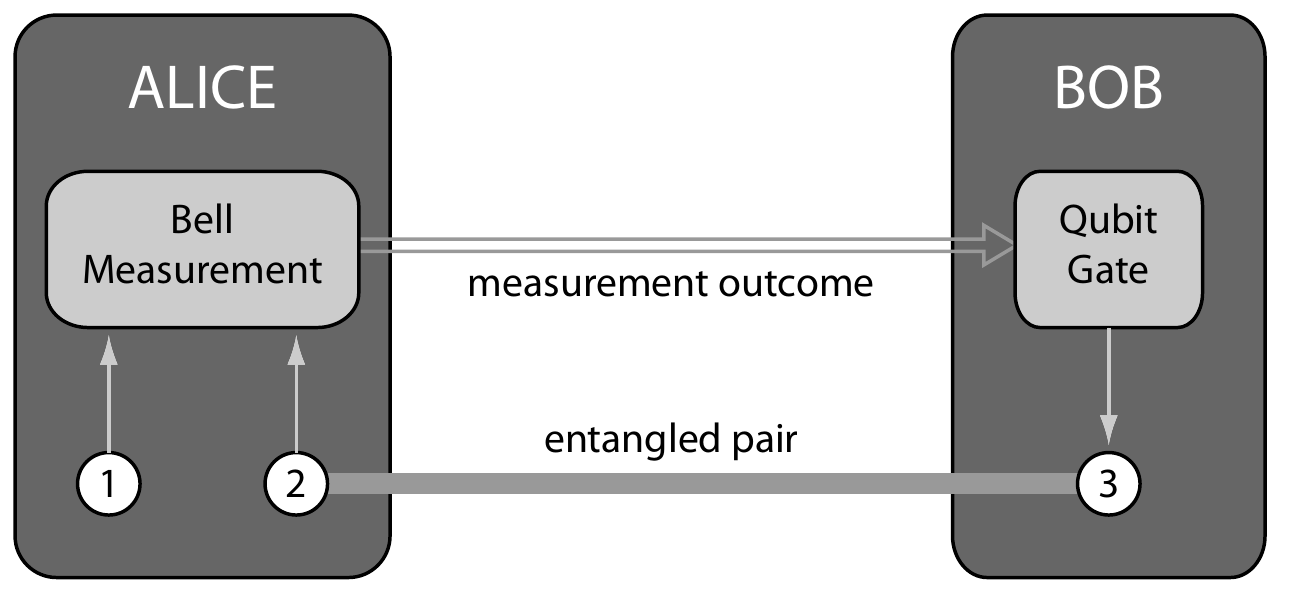}
\caption{Quantum teleportation. Alice takes a photon in an arbitrary quantum state (qubit 1) and one half of an entangled photon pair (qubit 2) and performs a Bell measurement. The outcome of the measurement determines the operation that Bob needs to perform on his photon (qubit 3). After teleportation is complete, the quantum state of qubit 1 has been transferred to qubit 3.} \label{fig:teleportation}
\end{minipage}
\end{center}
\end{figure}

In quantum teleportation, shown schematically in figure \ref{fig:teleportation}, Alice wishes to send an arbitrary quantum state of a photon to Bob. Rather than sending the state directly (which would be subject to photon loss), they first establish entangled photons pairs between each other. We denote the arbitrary quantum state by $\ket{\psi}_1$ and the shared entanglement is in the Bell state $\ket{\Phi^+}_{23}$, where the labels 1, 2, and 3 indicate the photons. Photons 1 and 2 are held by Alice, and photon 3 is held by Bob. The total state of the three photons is then given by 
\begin{align}
 \ket{\psi}_1 \ket{\Phi^+}_{23} = \left( \alpha\ket{H}_1 + \beta\ket{V}_1 \right) \frac{\ket{HH}_{23}+\ket{VV}_{23}}{\sqrt{2}} \equiv  \ket{\chi}_{123}\, ,
\end{align}
where $\alpha$ and $\beta$ are complex numbers obeying $\abs{\alpha}^2 + \abs{\beta}^2 = 1$. We can write this as
\begin{align}
 \ket{\chi}_{123} =  \frac{1}{\sqrt{2}} \left( \alpha\ket{HHH} + \alpha\ket{HVV} + \beta\ket{VHH} + \beta\ket{VVV}  \right) .
\end{align}
We suppressed the photon labels for brevity. 

Next, Alice performs a Bell measurement, in which her two photons are projected onto the Bell states. Writing the states $\ket{HH}$, $\ket{HV}$, $\ket{VH}$, and $\ket{VV}$ in the Bell basis and rearranging the terms, the state just before the Bell measurement is given by 
\begin{align}
 \ket{\chi}_{123} = & \frac12 \left[ \ket{\Phi^+} \left( \alpha \ket{H} + \beta\ket{V}\right) + \ket{\Phi^-} \left( \alpha \ket{H} - \beta\ket{V}\right) \right. \cr & + \left. \ket{\Psi^+} \left( \beta \ket{H} + \alpha\ket{V}\right) + \ket{\Psi^-} \left( \beta \ket{H} - \alpha\ket{V}\right) \right] .
\end{align}
The outcomes of Alice's Bell measurement indicate which state Bob's photon is in:
\begin{align}
  {\Phi^+}: & \quad \ket{\psi}_3 = \alpha \ket{H} + \beta\ket{V} \qquad  {\Psi^+}:  \quad \ket{\psi}_3 = \beta \ket{H} + \alpha\ket{V} , \cr
  {\Phi^-}: & \quad \ket{\psi}_3 = \alpha \ket{H} - \beta\ket{V} \qquad {\Psi^-}:  \quad \ket{\psi}_3 = \beta \ket{H} - \alpha\ket{V} . 
\end{align}
Bob does not know which of these states his photon is in until he receives a message from Alice telling him her measurement outcome. After receiving the message, Bob can apply a corrective operation (using half wave plates and quarter wave plates) to bring the quantum state of his photon back to the original $\alpha \ket{H} + \beta\ket{V}$. This completes the teleportation protocol. Quantum teleportation was demonstrated in 1997 and 1998 in various optical implementations \cite{bouwmeester97,boschi98,furusawa98}, and to various degrees of completeness \cite{kok00}. 

\begin{figure}[t!]
\begin{center}
\begin{minipage}{110mm}
\includegraphics[width=110mm]{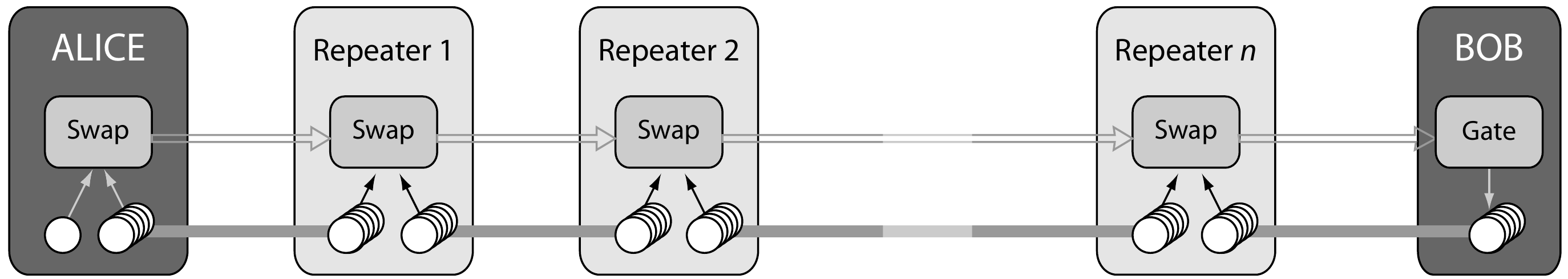}
\caption{A series of quantum repeaters can be used to extend the range of quantum communication. A quantum repeater teleports an input photon to a photon in the next repeater via the ``Swap'' operation. The entanglement between the repeaters is established beforehand and made near perfect using entanglement distillation. Bob receives a cumulative corrective instruction that is determined by all the Swap operations in the repeaters, and includes information which of his photons caries the teleported state.} \label{fig:repeater}
\end{minipage}
\end{center}
\end{figure}

A quantum repeater based on quantum teleportation works as follows (see figure \ref{fig:repeater}) \cite{briegel98}: Alice needs to send a polarised photon to Bob, who is too far away to send directly. Instead, she sends it to a repeater station somewhere between her and Bob. For now, let's assume that this station is close enough to Alice. The repeater station establishes shared entangled pairs with Bob, or another repeater (more on that later). This allows the repeater station to receive the incoming photon from Alice and teleport it to Bob using the shared entanglement (the ``Swap'' gate in figure \ref{fig:repeater}). For this to work, the Bell measurement must reveal whether the photon sent by Alice made it to the repeater station, and the entanglement between the repeater station and Bob must be (near) perfect. The repeater station then informs Bob what the corrective operation on his part of the entangled pair must be, and after making this correction Bob can measure his photon in the basis of his choice. Alice and Bob can now be far away from each other and still establish a secret key with a sufficiently high bit rate.

In the description above we have, of course, cheated! We magically assumed that the repeater station and Bob share near-perfect entangled photon pairs. This is far from trivial to establish. The photons must be created together, and at least one of them must travel the distance between the repeater station and Bob. That photon will inevitably incur losses of the same magnitude as the photon sent by Alice. The repeater station and Bob can share several pairs and try to find out which of the photons made it through. This is difficult, because it requires detecting a photon without destroying it (after all, we still need to use it in the teleportation protocol). Alternatively, we can perform entanglement distillation, which takes several imperfect entangled pairs and extracts one perfectly entangled pair. This is not easy either, because it requires entangling gates between photons. Finally, while all this processing is going on, the photons don't just sit there in the repeater station. They need to be actively stored in either an optical delay line or a quantum memory. If the distillation protocol requires communication between the repeater station and Bob, the length of time for which the photon needs to be stored is comparable to the time it takes light to travel from Bob to the repeater station. Any delay line memory would then incur the same amount of photon loss as the channel between the repeater station and Bob, and we're back to square one.

There are several architectures that attempt to circumvent these various difficulties, and one particular question of interest is what are the minimal requirements for a repeater to work? Does it need two-way communication between stations or can we construct a protocol that requires only one-way communication as shown in  figure \ref{fig:repeater}? Does the repeater require memories that last as long (or longer) than the flight time of the photons between repeater stations? A lot of progress is being made on these questions, and this is currently an active area of research \cite{briegel98,kok02,jiang09,zwerger12,azuma15,reiserer15}.

Finally, we note that while anti-bunching and the no-cloning theorem are sufficient for the design of quantum key distribution, the implementation of this protocol over long distances will most likely require the use of polarisation entanglement. This lifts the construction of repeaters into a new realm of difficulty over direct quantum communication.

\section{Quantum metrology and imaging}\label{sec:metro}
\begin{figure}[t!]
\begin{center}
\begin{minipage}{110mm}
\includegraphics[width=90mm]{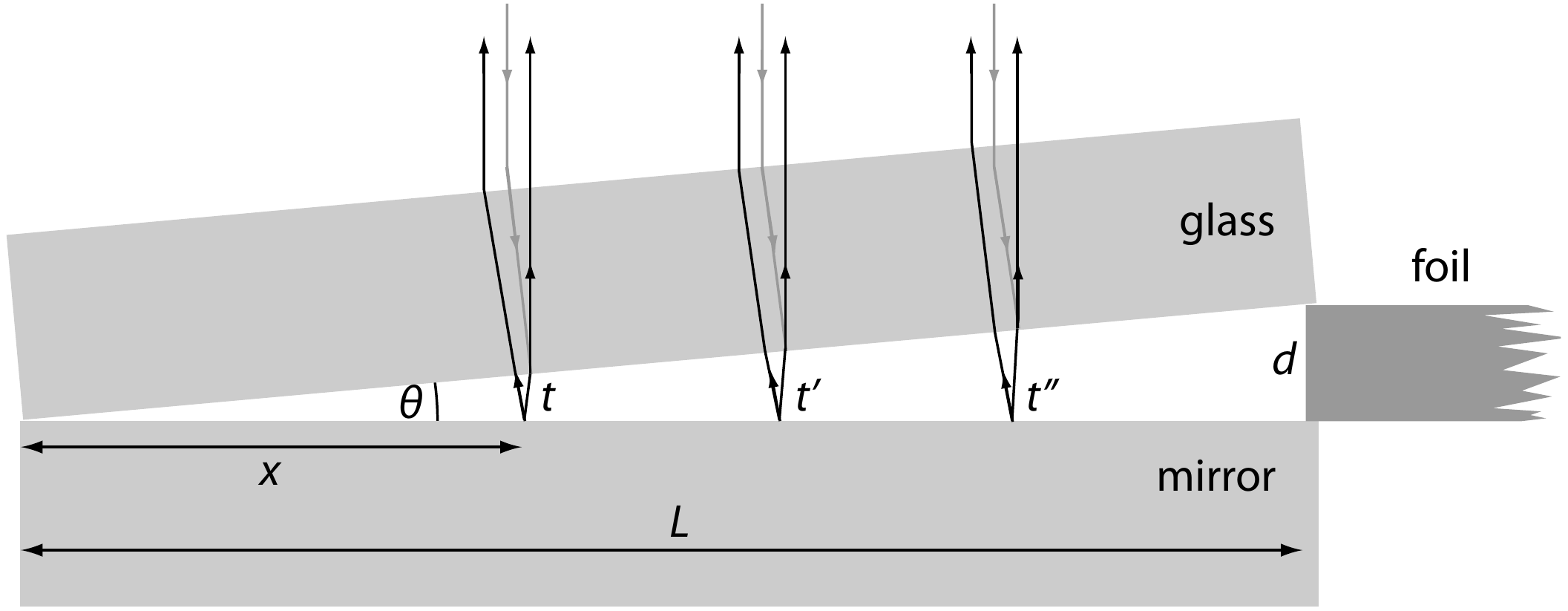}
\caption{Light can be used to measure the thickness $d$ of a piece of foil by counting the number of interference fringes along a distance $L$ as seen from above. The precision in the measurement of $d$ can be estimated as $\Delta d \sim \frac12 \lambda$, where $\lambda$ is the wavelength of the light. This is much more precise than a mechanical micrometer, which has a precision of about 0.05 mm.} \label{fig:umm}
\end{minipage}
\end{center}
\end{figure}

Light is also extremely useful in metrological applications. As a simple example, consider the measurement of the thickness of a thin piece of foil. A traditional mechanical micrometer has a precision of about 0.05~mm, which is not good enough to measure foils that are much thinner\footnote{Metal foils a few micrometres thick are readily available.}. To obtain the required precision, we can use an optical interferometric micrometer shown in figure~\ref{fig:umm}, the principle of which is identical to that of Newton's rings. The foil is placed at the end of a mirror and holds up a plate of glass. The reflected light will consist of two contributions, namely the light that is reflected off the inside surface of the glass plate, and the light that is reflected off the mirror. Whether constructive or destructive interference occurs at the outgoing light depends on the path difference $t$ between the two contributions. In figure~\ref{fig:umm}, we show this effect at three different positions along the mirror.  

We find destructive interference between the two reflected light waves when the path length $t$ is a half-integer multiple of the wavelength $\lambda$:
\begin{align}
 t = \left( m + \frac12 \right) \lambda = x \tan\theta = \frac{xd}{L}\, ,
\end{align}
where $m$ is an integer, $\theta$ is the angle between the glass plate and the mirror, $L$ is the distance between the pivot point and the foil, and $x$ is the distance from the pivot point to the dark fringe. Adjacent fringes are a distance $\Delta x$ apart, with 
\begin{align}
 \Delta x = \frac{\lambda L}{2d} \qquad\text{or}\qquad d = \frac{\lambda L}{2\Delta x} \equiv \frac12 \lambda L \sigma \, ,
\end{align}
where $\sigma$ is the number of fringes per unit length. If we know $\lambda$ and $L$, we can count the fringes to obtain $\sigma$, which in turn gives us a value for the foil thickness $d$.

The precision of this measurement can be estimated by noting that the number of fringes per unit length $\sigma$ can be counted very accurately, as long as there is at least one fringe across the length $L$, or $\Delta\sigma \sim 1/L$. Using the error propagation formula we calculate $\Delta d$ from this estimate as follows:
\begin{align}
 \Delta d = \frac{\Delta\sigma}{|{\rm d}\sigma/{\rm d}d|} = \frac{\lambda}{2} L\Delta\sigma \sim  \frac{\lambda}{2}\, .
\end{align}
We see that this device can reach a precision of a few hundred nanometers, which is two to three orders of magnitude better than the mechanical micrometer. The above example is using only classical light. Can we improve on this technique by using quantum light? In order to answer this question we will have to go back to what makes quantum light different from classical light.

\bigskip

\noindent
We can in principle improve the precision of the optical micrometer by using a lot of light: If we measure the intensity along the $x$-direction with very high precision we can detect any variation in intensity, even if that amounts to much fewer than one fringe per length $L$. There is a catch, however, since the intensity itself is noisy. A (transversely) coherent state of light with on average $\braket{n}$ photons has intensity fluctuations that are proportional to $\sqrt{\braket{n}}$. The signal to noise ratio (SNR) is then $\braket{n}/\sqrt{\braket{n}} = \sqrt{\braket{n}}$. The more photons, the higher the SNR, and the higher the SNR, the more precisely the intensity curve in the $x$-direction can be measured (inversely proportional to the SNR). The number of photons is therefore a \emph{resource} for measuring the foil thickness $d$: the more you have of them, the better you can estimate $d$. The precision of $d$ will scale according to 
\begin{align}\label{eq:sql}
 \Delta d \propto \frac{1}{\sqrt{\braket{n}}}\, .
\end{align}
This is called the \emph{shot noise limit}, or the \emph{standard quantum limit} (SQL). It originates in the natural intensity fluctuations of light. The $1/\sqrt{\braket{n}}$ behaviour is specific to the type of light, which in this case is classical coherent light. We therefore sometimes refer to this precision as the classical precision limit. 

If there is a way to suppress these fluctuations in the intensity we may be able to increase the precision $\Delta d$. How would this work? To answer this, note that the photons arrive in the detector completely independently of each other, which means that they will cluster randomly at each pixel in the detector according to the Poisson distribution in Eq.~(\ref{eq:poisson}). To remove this randomness we need a ``conspiracy'' between the photons in the form of transverse anti-bunching (e.g., see figure~\ref{fig:hbt}, where $\tau$ may now denote the distance between the detected positions). If the photons arrive nicely spaced out, the intensity fluctuations at each pixel will be suppressed. The periodic structure of the fringes will then become clear much quicker, and an accurate count of the fringes can be performed with fewer photons \cite{boyer15}. 

However, there is a limit to the precision gain that can be obtained this way. No matter how evenly spaced, we still need an appreciable number of photons to reveal the fringes. The ultimate precision in $d$ can be calculated as
 \begin{align}
 \Delta d \propto \frac{1}{\braket{n}}\, .
\end{align}
This is called the \emph{Heisenberg limit} \cite{holland93}. Since reaching this limit requires anti-bunching (which in this case will require some form of entanglement between the photons) this is a truly quantum mechanical precision scaling without a classical implementation. 

More generally, the ultimate precision allowed by quantum mechanics of the measurement of a parameter $\theta$ is given by the expectation value of the operator $K$ that drives the changes in that parameter. These operators are called \emph{generators}. For example, the generator for changes in time is the Hamiltonian, the generator for translations in space is the momentum operator, and the generator for phase changes is the number operator. The unitary evolution that imparts the parameter $\theta$ onto the quantum state is given by $U(\theta) = \exp(-i K \theta/\hbar)$. The ultimate precision is then written as a \emph{lower bound} on the root mean square error of $\theta$ \cite{braunstein96}:
 \begin{align}\label{eq:hl}
 \Delta\theta \geq \frac{\hbar}{2} \frac{1}{\Delta{K}}\, .
\end{align}
For optimal states, the quantum mechanical operator variance $(\Delta K)^2$ is bounded by the (squared) expectation value $\braket{K}^2$, and $\braket{K}$ is the proper definition of the resource (e.g., the amount) that allows us to increase the precision of the measurement of $\theta$ \cite{zwierz10}. The expectation value $\braket{K}$ is often much easier to estimate than the variance $(\Delta K)^2$.

From Eq.~(\ref{eq:hl}) it is clear why this limit is called the Heisenberg limit: if $\theta$ is the position $x$ of a particle and $K$ is its momentum $p$, then the inequality becomes 
 \begin{align}
 \Delta x\, \Delta p \geq \frac{\hbar}{2} \, ,
\end{align}
which you will recognise as Heisenberg's uncertainty relation for position and momentum. Eq.~(\ref{eq:hl}) is more general than the traditional Heisenberg-Robertson relation that is derived only for (non-commuting) observables, since it is valid also for physical quantities that do not have an associated quantum operator, such as time, phase and rotation angle.

It is often argued that entanglement is a prerequisite for reaching the Heisenberg limit \cite{giovannetti06,giovannetti11}. While this is certainly true in the context of estimation procedures involving many distinguishable particles, the situation in optics---where photons may be indistinguishable---is a little more subtle. Since the Heisenberg limit in Eq.~(\ref{eq:hl}) depends on $\Delta{K}$, we can in principle construct a quantum state on a single optical mode that maximises $\Delta{K}$. For example, if we wish to measure an optical phase $\varphi$, the relevant generator is the number operator. The state with a maximal variance (and bounded maximum number states) is given by 
\begin{align}\label{eq:0N}
 \ket{\psi(0)} = \frac{\ket{0} + \ket{N}}{\sqrt{2}}\, ,
\end{align}
with $\ket{0}$ the state of no photons, and $\ket{N}$ the state of $N$ photons in the optical mode. A phase shift on the optical mode then leads to the transformation
\begin{align}
  \ket{\psi(0)} = \frac{\ket{0} + \ket{N}}{\sqrt{2}} \quad\to\quad \frac{\ket{0} + {\rm e}^{iN\varphi}\ket{N}}{\sqrt{2}} = \ket{\psi(\varphi)}\, .
\end{align}
Measuring the observable $X_N = \ket{N}\bra{0} - \ket{0}\bra{N}$ will then give a precision \cite{koklovett10}
\begin{align}
 \Delta\varphi = \frac{\pi}{2} \frac{1}{N}\, .
\end{align}
There is no entanglement in a single optical mode, but we still attain the Heisenberg limit. In any real estimation procedure, however, the use of entanglement can help overcome practical difficulties such as creating the superposition in Eq.~(\ref{eq:0N}) or implementing the observable $X_N$. For example, instead of the state in  Eq.~(\ref{eq:0N}), we may want to create the so-called {\sc noon} state \cite{bollinger96,lee02}
\begin{align}\label{eq:noon}
 \ket{\psi(0)} = \frac{\ket{N,0} + \ket{0,N}}{\sqrt{2}}\, ,
\end{align}
which is a two-mode state in which all the photons are in one mode (but it is undetermined which mode). A phase shift in one of the modes will then induce the same relative phase factor ${\rm e}^{iN\varphi}$ as in Eq.~(\ref{eq:0N}). However, since this is not a superposition of different photon numbers but rather a superposition of the distribution of $N$ photons, it is conceptually easier to see how this can be made in practice \cite{mitchell04,walther04}. 

Still, creating {\sc noon} states is extraordinarily difficult, and they are extremely sensitive to decoherence. More promising is the use of \emph{sqeezed} light. This is another type of quantum mechanical light that has no classical analog. Instead of considering the photon number, which are the eigenvalues of the operator $\hat{n} = \hat{a}^\dagger \hat{a}$, we may look at the \emph{quadrature} operators
\begin{align}
 \hat{X} = \frac{\hat{a}+\hat{a}^\dagger}{2} \qquad\text{and}\qquad  \hat{Y} = -i \frac{\hat{a}-\hat{a}^\dagger}{2}\, .
\end{align}
The commutation relation between $\hat{X}$ and $\hat{Y}$ is $[\hat{X},\hat{Y}] = i$, which means that they obey the uncertainty relation 
\begin{align}\label{eq:quad}
 \Delta X \, \Delta Y \geq \frac12\, .
\end{align}
To get an idea what these operators mean, remember that the creation and annihilation operators are mathematically identical to the ladder operators of the simple harmonic oscillator. By analogy, $\hat{X}$ and $\hat{Y}$ behave as the position and momentum of the simple harmonic oscillator. Measuring $\hat{X}$ would then be equivalent to measuring the amplitude of the oscillator, while measuring $\hat{Y}$ would be equivalent to measuring the momentum of the oscillator. In the language of waves these are called quadratures.

A classical coherent state of light is a minimum uncertainty state, in the sense that Eq.~(\ref{eq:quad}) becomes an inequality. Not only that, the two variances are also equal:
\begin{align}
 (\Delta X)^2 = (\Delta Y)^2 = \frac12\, .
\end{align}
Quantum mechanically, we can reduce the variance in one quadrature at the expense of the other while still obeying Eq.~(\ref{eq:quad}). We can write this as 
\begin{align}
 (\Delta X)^2 = \frac{{\rm e}^{-2r}}{2}  \qquad\text{and}\qquad  (\Delta Y)^2 =  \frac{{\rm e}^{2r}}{2}  \, ,
\end{align}
where $r$ is called the squeezing parameter. Using these types of light allows us to achieve a measurement precision in $\varphi$ of
\begin{align}
 \Delta\varphi \geq \frac{{\rm e}^{-r}}{\sqrt{M \braket{n}}}\, ,
\end{align}
where $\braket{n}$ is the average number of photons in the light probe, and $M$ is the number of times the experiment is repeated \cite{pezze08}. The advantage of this approach is that it does not require too exotic quantum states such as the {\sc noon} state, and the measurement can be achieved by ordinary homodyne detection. This method is proposed as part of Advanced LIGO for the measurements of gravitational waves \cite{caves81,aligo13}

\begin{figure}[t!]
\begin{center}
\begin{minipage}{110mm}
\includegraphics[height=60mm]{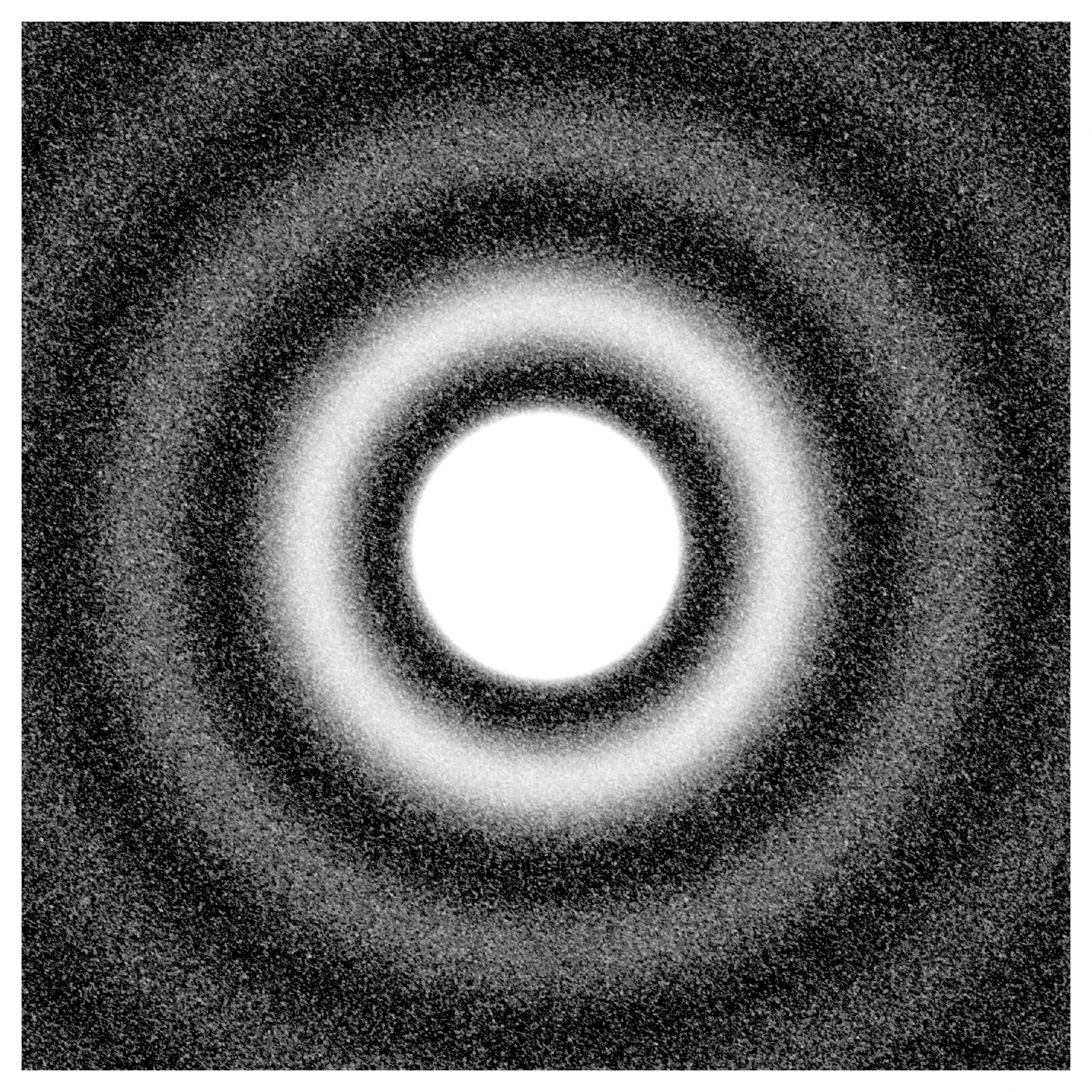}
\caption{The image of a small circular source exhibits diffraction fringes, and has a smeared out character that makes it difficult to find the exact position and radius of the source. If the image was perfectly smooth we would be able to characterise its parameters from Fourier analysis, but the intensity profile has a fair amount of noise that leads to an error in the estimation of the source dimensions.} \label{fig:aperture}
\end{minipage}
\end{center}
\end{figure}

Finally, the same transverse anti-bunching effect that was used to increase the precision of the optical micrometer can be employed to improve the resolution in imaging. Suppose that we wish to image an object that we know is circular (for example a star or a small aperture). We can use a telescope or a microscope and obtain an image like the one shown in figure~\ref{fig:aperture}. We may infer the radius (or, more precisely, the opening angle) of the sources, as well as the position of the source, by matching the intensity of the light in the imaging plane with a theoretical model of the image (related to the Fourier transform of the source geometry). Given a perfectly smooth intensity profile in the imaging plane we can find the position and radius with arbitrarily high precision. However, as before the intensity of the light in the image plane fluctuates, and this will create a degree of uncertainty in the fitting of the theoretical curve to the data \cite{perezdelgado12,pearce15}. If the fluctuations can be suppressed via transverse anti-bunching, the precision of the estimates of the position and radius can achieve the Heisenberg limit \cite{boyer15}.

%We see that imaging is at heart a form of parameter estimation using many variables. 

% multi-parameter estimation and nuisance parameters

\section{Quantum computing}\label{sec:comp}
The last, and arguably most challenging information processing task with single photon qubits is quantum computing. Quantum computers have very stringent noise requirements, which linear optics can in principle meet. The major challenge, however, is in the generation of entanglement. To make matters worse, not all entanglement in created equal.

\subsection{Entanglement generation between photons}
One of the key ingredients in quantum information processing is quantum entanglement. For example, in Section~\ref{sec:comm} we used the maximally entangled Bell states as resources for quantum teleportation. For quantum computing, entanglement is also a key resource. No exponential speed-up can be achieved without it. However, when we deal with photons as our information carriers, we must distinguish between different types of entanglement.

Consider a single photon impinging on a 50:50 beam splitter. There are two input and two output modes for a simple beam splitter, and we can write the mode transformations as 
\begin{align}
 \hat{a}_1 \to \frac{\hat{b}_1+\hat{b}_2}{\sqrt{2}} \qquad\text{and}\qquad \hat{a}_2 \to \frac{\hat{b}_1-\hat{b}_2}{\sqrt{2}} \, ,
\end{align}
where $\hat{a}_1$ and $\hat{a}_2$ are the annihilation operators for the input modes, and $\hat{b}_1$ and $\hat{b}_2$ are the annihilation operators for the output modes. A single photon entering the beam splitter in mode 1 can then be written as a quantum state transformation
\begin{align}
 \ket{1,0}_{12} \to \frac{ \ket{1,0}_{12}+ \ket{0,1}_{12}}{\sqrt{2}}\, .
\end{align}
This state is entangled. It can in principle be used to violate a Bell inequality (even though it would be difficult to implement in practice). The entanglement is between the spatial degree of freedom (mode 1 or 2), and the photon number degree of freedom (0 or 1 photons). In general, quantum states that are not thermal or classical coherent states become entangled when they interact with beam splitters \cite{koklovett10}.

Unfortunately, this type of entanglement is of limited use for quantum computation. To see this, consider a Bell state required for quantum computing:
\begin{align}
 \ket{\Phi^+}_{12} = \frac{ \ket{H,H}_{12} + \ket{V,V}_{12}}{\sqrt{2}}\, .
\end{align}
This is a state of two photons with polarisation degree of freedom ($H$ and $V$) in two spatial modes (1 and 2). We can write this in terms of creation operators acting on the vacuum as
\begin{align}
 \ket{\Phi^+}_{12} = \frac{1}{\sqrt{2}}\left(  \hat{a}_{1,H}^\dagger \hat{a}_{2,H}^\dagger + \hat{a}_{1,V}^\dagger \hat{a}_{2,V}^\dagger \right) \ket{0}\, .
\end{align}
Suppose that we wish to create this state from the separable input state $\ket{H,H}$. The mode transformation that must be implemented is then 
\begin{align}\label{eq:insep}
 \hat{a}_{1,H}^\dagger \hat{a}_{2,H}^\dagger \to  \frac{1}{\sqrt{2}}\left(  \hat{b}_{1,H}^\dagger \hat{b}_{2,H}^\dagger + \hat{b}_{1,V}^\dagger \hat{b}_{2,V}^\dagger \right)\, .
\end{align}
Linear optics is \emph{linear} in the mode transformations, which means that each input mode operator transforms into a linear combination of the output mode operators. In other words, 
\begin{align}\label{eq:sep}
 \hat{a}_{1,H}^\dagger \to \sum_{j,s} U_{1j,Hs} \hat{b}_{j,s}\, ,
\end{align}
where $U_{1j,Hs}$ are the elements of a unitary matrix\footnote{There is a one-to-one relation between unitary matrices and linear optical networks that consist of beam splitters, phase shifters and polarisation rotations \cite{reck94}.}. Each mode operator is replaced with a sum over mode operators. However, the substitution rule of Eq.~(\ref{eq:sep}) applied to the left-hand side of Eq.~(\ref{eq:insep}) can never produce the right-hand side of Eq.~(\ref{eq:insep}) because the left-hand side is separable into a product of two mode operators, whereas the right-hand side is not. Therefore, linear optics alone cannot be used to create the necessary entanglement for quantum computing.

One potential way around this problem is to use an induced photon-photon interaction, for example using a Kerr nonlinearity. Such a nonlinearity imparts a phase shift on one optical mode that is proportional to the intensity in another mode. At the single-photon level this can act as a coherent switch. Unfortunately, Kerr nonlinearities are inherently noisy and cannot be used for single-photon quantum gates \cite{shapiro06}. The question is then whether we can use photonic qubits for quantum computing.

\subsection{The Knill-Laflamme-Milburn protocol}
\begin{figure}[t!]
\begin{center}
\begin{minipage}{110mm}
\includegraphics[height=25mm]{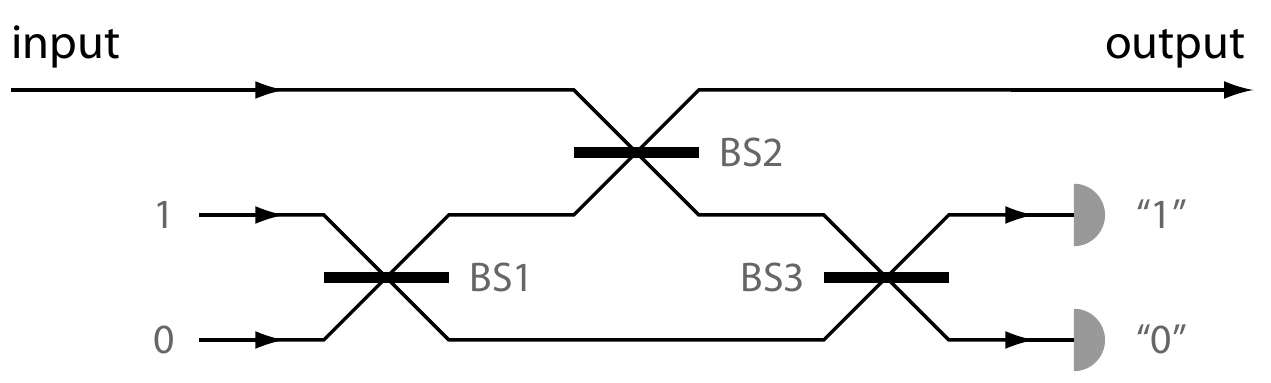}
\caption{A linear optical network that implements a nonlinear phase shift (an NS gate) using only beam splitters and projective measurements. Instead of each photon in the input mode accumulating a $-1$ phase shift, only the two-photon component picks up the $-1$ phase shift, leaving the one-photon component unaffected. The implementation requires one ancilla photon in the middle mode, and a detection signature of one photon and no photons in the two detectors in the output. The beam splitters BS1, BS2 and BS3 are not 50:50, but have specially chosen transmission coefficients.} \label{fig:ns}
\end{minipage}
\end{center}
\end{figure}

The problem was solved in 2000 by Knill, Laflamme and Milburn, in what was to become one of the classic papers in quantum information processing \cite{klm01}. Instead of a medium-induced photon-photon interaction, the required nonlinearity of the mode transformations is provided by projective measurements. In addition to the photons that are part of a computation, we may send extra \emph{ancilla} photons through a linear optical network of beam splitters and phase shifters. This gives us the freedom to detect photons in very specific output modes, as shown in the example of the nonlinear phase shift circuit in figure~\ref{fig:ns}, also known as the NS gate. Since the  number of added ancilla photons is the same as the number of detected photons, the photon number in the input mode does not change once it has passed through the network.  

Of course, it is not guaranteed that the two detectors will detect one and zero photons, respectively. If it was, there would be no need for detection. This implies that the circuit in figure~\ref{fig:ns} succeeds only part of the time (in this case, the success probability of the gate is one quarter). This is no good for quantum computation, in which all the circuits must be successful simultaneously. To overcome this problem, Knill, Laflamme and Milburn employed quantum teleportation: Instead of trying to apply the probabilistic gate directly to the quantum information carrying qubits (which cannot be copied and must therefore be handled with care), the gate is applied to one half of an entangled pair. If the gate is successful, the now modified entangled pair is used as the entanglement resource in quantum teleportation of the information carrying qubit. The teleported qubit emerges with the gate applied to it. Knill, Laflamme and Milburn found a way to make the teleportation procedure nearly deterministic, which means that the probabilistic gate can now be applied deterministically to the qubit, and linear optical quantum computing was in principle possible. 

\begin{figure}[t!]
\begin{center}
\begin{minipage}{110mm}
\includegraphics[height=30mm]{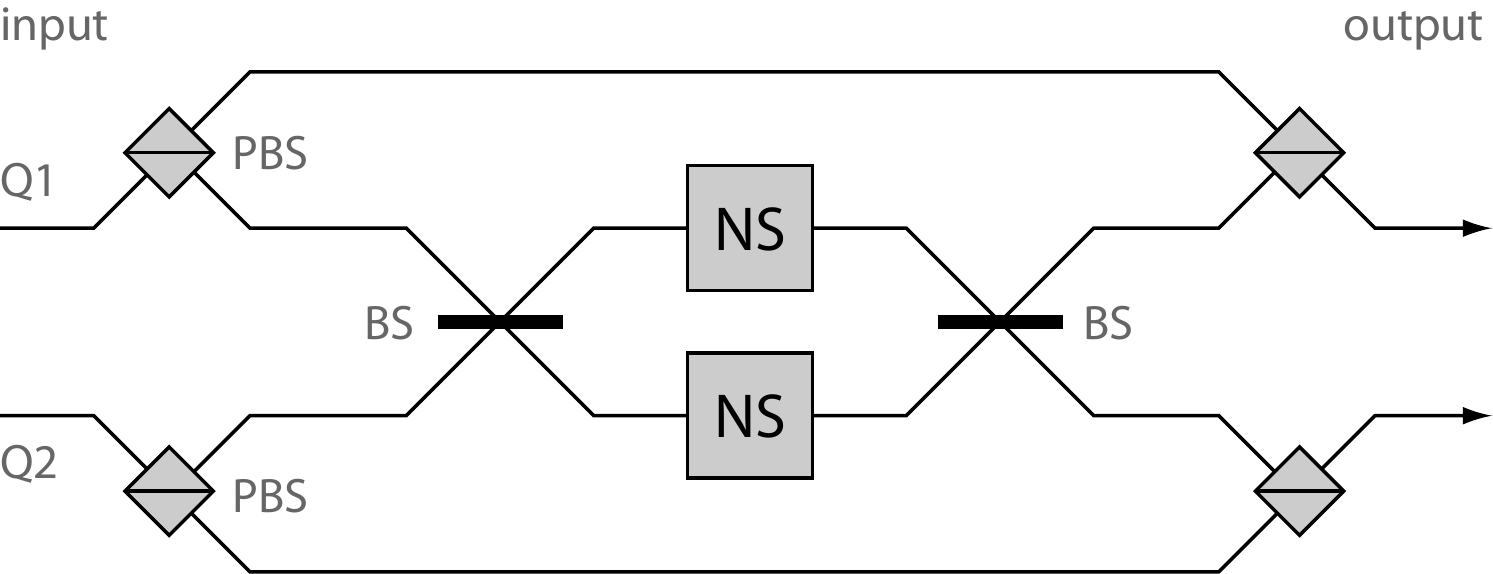}
\caption{The controlled-$\sigma_z$ (CZ) gate. Two photonic qubits, Q1 and Q2, enter the interferometer. The modes corresponding to qubit value $\ket{1}$ are sent into a 50:50 beam splitter, the output of which are subject to a nonlinear phase shift (NS). If there is a photon in each input mode of the beam splitter, the Hong-Ou-Mandel effect guarantees that the two photons will either go both through the top NS gate or through the bottom NS gate. Consequently, these photons will pick up a $-1$ phase. The second 50:50 beam splitter will separate the two photons again into one photon in each output mode of the beam splitter.} \label{fig:cz}
\end{minipage}
\end{center}
\end{figure}

Once we can create an NS gate, we can use the Hong-Ou-Mandel effect to create controlled Pauli $\sigma_z$ gates, or CZ gate. These are the two-qubit gates that can create the entanglement necessary for quantum computing. In terms of qubits, the CZ gate operates as follows on the two-qubit states:
\begin{align}\label{eq:cz}
 U_{CZ} \ket{00} = \ket{00}\, , & \quad
 U_{CZ} \ket{01} = \ket{01}\, , \cr
 U_{CZ} \ket{10} = \ket{10}\, , & \quad
 U_{CZ} \ket{11} = -\ket{11}\, .
\end{align}
In other words, when both qubits are in the $\ket{1}$ state, the gate applies a phase shift ${\rm e}^{i\pi} = -1$. To see how this gate can be implemented with two NS gates and the Hong-Ou-Mandel effect, consider the circuit in figure~\ref{fig:cz}. We can arrange the two incoming qubits Q1 and Q2 in such a way that the $\ket{0}$ states for each qubit---i.e., horizontal polarisation---are mapped onto the top and bottom modes that propagate freely. The $\ket{1}$ states for each qubit are the vertically polarised photons, and will be reflected in the polarising beam splitters (PBS). The photons will combine in the first 50:50 beam splitter. If an input state $\ket{V,V}$ enters this circuit, both photons will meet at the first beam splitter and experience to Hong-Ou-Mandel effect. This means that both photons will exit the beam splitter in a quantum superposition of both photons in the top mode and both photons in the bottom mode. The NS gate, if successful, will then impart a $-1$ phase shift on the two-photon state. The second beam splitter will apply the Hong-Ou-Mandel effect in reverse, such that 
\begin{align}
 \ket{V,V} ~\to~ \frac{\ket{2V,0}-\ket{0,2V}}{\sqrt{2}} ~\to~  \frac{-\ket{2V,0}+\ket{0,2V}}{\sqrt{2}} ~\to~ -\ket{V,V}\, .
\end{align}
If only one photon enters the first beam splitter, for example because Q1 is in the logical state $\ket{1}$ and Q2 is in the state $\ket{0}$, there will only be one photon going through the NS gates, and there will be no phase shift. Similarly, when both photons are in the top and bottom mode, no photons travel through the NS gates and no phase shift is imported on the quantum state. The result is that the circuit in figure~\ref{fig:cz} implements the gate in Eq.~(\ref{eq:cz}). The gate is successful when both NS gates are successful, and the total success probability is therefore $p_{\rm CZ} = (\frac14)^2 = 1/16$. The gate can be applied to qubits in the computation using the teleportation trick described above.

It is important for the operation of the CZ gate that the Hong-Ou-Mandel effect works perfectly. This means that the two photons must be indistinguishable in every respect, including frequency, polarisation and mode shape. Imperfections in the photon source, the beam splitters, or the NS gate will create faulty gates that can ruin the computation.

\subsection{Measurement-based quantum computing}
The Knill-Laflamme-Milburn protocol is a type of measurement-based quantum computing, in which the computations are induced by measurements and feed-forward processing of the measurement outcomes. As a practical scheme, however, it has many downsides: a single entangling gate needs tens to hundreds of thousands of ancilla photons; the detectors must be nearly perfectly efficient and be able to tell the difference between 0, 1 and 2 photons; all the photons must be identical to an extremely high degree; and the feed-forward procedure requires high-quality, low-loss optical switches. In addition, while the feed-forward takes place, the photons must be stored in a quantum memory.

\begin{figure}[t!]
\begin{center}
\begin{minipage}{110mm}
\includegraphics[width=110mm]{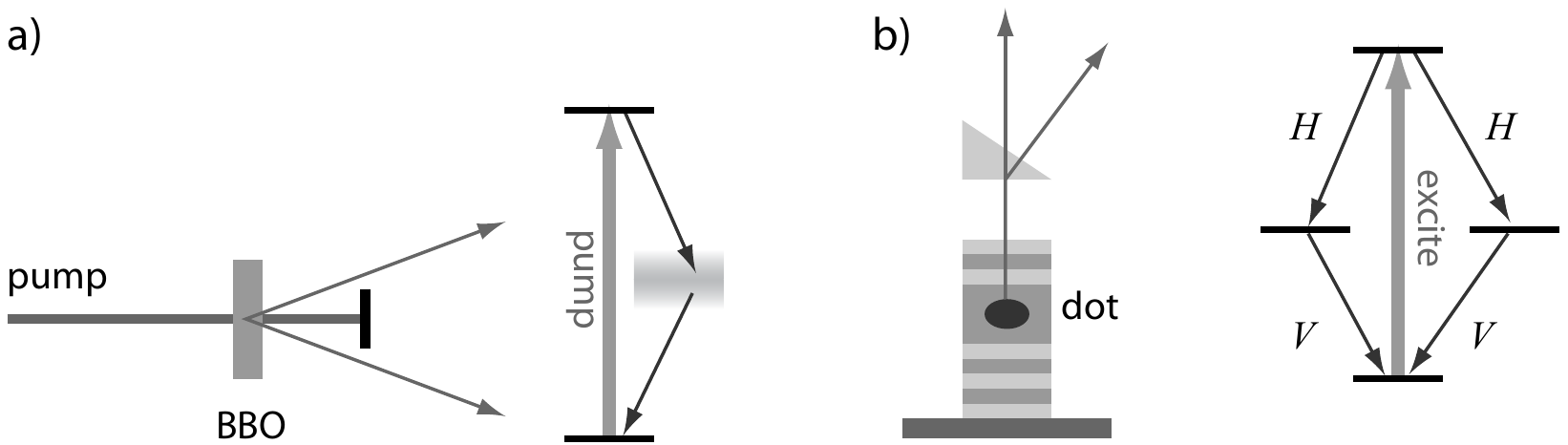}
\caption{Creating two-photon entangled states. a) Spontaneous parametric down-conversion creates photon pairs probabilistically when one or more pump photons break down into two entangled photons whose frequencies sum to the pump frequency. This can be achieved for example in a BBO crystal. Two BBO crystals back-to-back rotated ninety degrees with respect to each other will create polarisation-entangled photon pairs. b) A single quantum dot in a Bragg stack can be excited and decay along two different paths, creating polarisation-entangled photons on demand. However, while this gives in principle superior performance over SPDC, the fabrication challenges are significant and this approach is currently still in the research stage.} \label{fig:spdc}
\end{minipage}
\end{center}
\end{figure}

The first problem, the resource count, can be mitigated if instead of single photon ancilla states we use entangled photons from the start. This requires a reliable source of photons in one of the Bell states (it does not really matter which one). Traditionally, entangled photon pairs have been generated using a process called Spontaneous Parametric Down-Conversion (SPDC), in which a high energy  laser pumps a nonlinear crystal. The photons of the laser have a very small probability of ``breaking up'' in the crystal into two photons of lower energy. Depending on the arrangement, these two photons can be created in an entangled polarisation state $(\ket{H,H}+\ket{V,V})/\sqrt{2}$ \cite{kwiat95}. Alternatively, we can engineer quantum dot structures that create entangled photons on demand \cite{stevenson06} as shown in figure~\ref{fig:spdc}. The dot can be placed in a Bragg stack that sends photons in the vertical direction, and a prism separates the different frequency components. While SPDC is clean and straightforward to implement, the rate of photon pair production is extremely low, and occasionally two or more pairs are created. The quantum dot approach would therefore be preferable, but it is currently still in the research stage.

Assuming that we have a reliable two-photon source we can design a new architecture for linear optical quantum computing that requires significantly fewer resources. The key is still to use gate teleportation, but instead of the complicated states required by the Knill-Laflamme-Milburn protocol we create conceptually (and practically) simpler cluster states. Consider two polarised photons in the (unnormalised) entangled state
\begin{align}\nonumber
 \ket{H,H} + \ket{H,V} + \ket{V,H} - \ket{V,V} .
\end{align}
We can measure the first photon in a special basis ``$\pm\alpha$'' with eigenstates
\begin{align}
 \ket{+\alpha} \equiv \frac{\ket{H} + {\rm e}^{i\alpha}\ket{V}}{\sqrt{2}} \qquad\text{and}\qquad \ket{-\alpha} \equiv \frac{\ket{H} - {\rm e}^{i\alpha}\ket{V}}{\sqrt{2}}\, .
\end{align}
After finding, say the measurement outcome $+\alpha$ the quantum state of the remaining photon is 
\begin{align}
 \ket{\psi_{\rm out}} = \frac{1+{\rm e}^{-i\alpha}}{2} \ket{H} + \frac{1-{\rm e}^{-i\alpha}}{2} \ket{V} = H U_Z(\alpha) \left( \frac{\ket{H}+\ket{V}}{\sqrt{2}} \right)\, ,
\end{align}
where the last equality is true up to an unobservable global phase, and $U_Z(\alpha) = \exp(-i\alpha\sigma_z/2)$ is a rotation generated by the Pauli $\sigma_z$ operator. In other words, measuring the first photon in the special basis ``$\pm\alpha$'' produces a unitary gate $H U_Z(\alpha)$ on the second photon. We can daisy-chain this process by using a four-photon entangled state and measuring the first three photons in special bases defined by successive angles $\alpha$, $\beta$, and $\gamma$. The resulting operation on the final photon is the unitary gate 
\begin{align}
 U(\alpha,\beta,\gamma) = H U_Z(\gamma) H U_Z(\beta) H U_Z(\alpha) = H U_Z(\gamma) U_X(\beta) U_Z(\alpha)\, ,
\end{align}
where we have used that $H U_Z H = U_X$, a rotation generated by the Pauli $\sigma_x$ operator. Such a gate can implement any single qubit operation given judiciously chosen values of $\alpha$, $\beta$, and $\gamma$. Depending on the (probabilistic) measurement outcome ($\pm\alpha$), the subsequent measurement angle must be chosen as $\pm\beta$, and the measurement outcome $\pm\beta$ determines the angle $\pm\gamma$. This forward dependence of the measurement angles creates a definite direction of the computation.

The four-photon state can be graphically represented as a linear (one-dimensional) graph, in which the nodes denote the photons and the edges denote entanglement created by CZ gates between the photons:
\begin{align}\label{eq:1Dcluster}
 \ket{\psi_{\rm 1D}} = \vcenter{\hbox{\includegraphics[width=40mm]{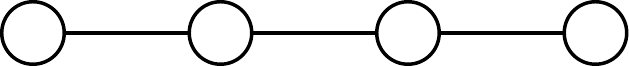}}}
\end{align}
We can create two-dimensional graphs in which the vertical entanglement connections represent entangling gates:
\begin{align}\label{eq:2Dcluster}
 \ket{\psi_{\rm 2D}} = \vcenter{\hbox{\includegraphics[width=40mm]{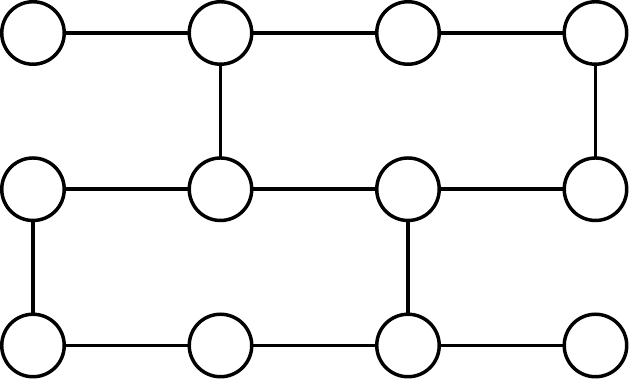}}}
\end{align}
These structures can be mapped onto any quantum computational circuit and are therefore universal for quantum computation. The entangled states in Eqs.~(\ref{eq:1Dcluster}) and (\ref{eq:2Dcluster}) are called \emph{cluster states}, and the method is called  \emph{one-way} quantum computation \cite{raussendorf01}. Since photon measurements can in principle be carried out efficiently, the challenge is to create the required cluster states.

\begin{figure}[t!]
\begin{center}
\begin{minipage}{110mm}
\includegraphics[width=110mm]{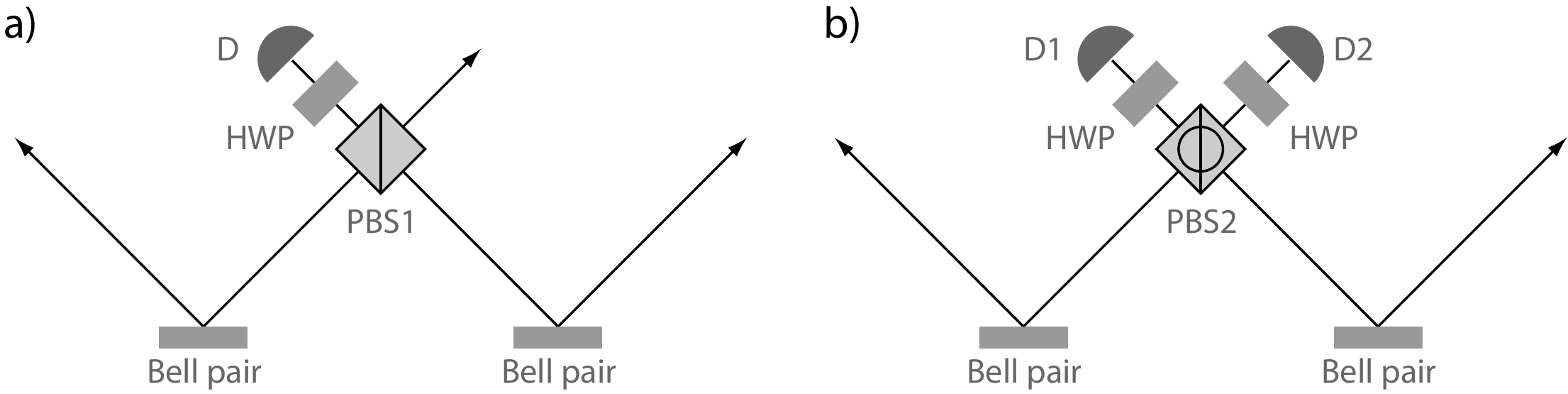}
\caption{Fusion gates for creating cluster states. a) Type-I fusion allows us to take two Bell pairs and create an entangled three-photon state by measuring a single photon in detector D after mixing on a linear polarising beam splitter (PBS1) and a half wave plate (HWP). b) Type-II fusion is a form of entanglement swapping that uses a circular polarising beam splitter (PBS2) and two detection events at D1 and D2. Here, the fusion gate is applied to two Bell pairs, which results in another pair. In larger systems the type-II fusion gate can be used to create bigger cluster states without the noise drawbacks of the type-I fusion gate.} \label{fig:fusion}
\end{minipage}
\end{center}
\end{figure}

A particularly promising way to create large cluster states is to use so-called \emph{fusion gates}, shown in figure~\ref{fig:fusion} \cite{browne05}. The entanglement is created by a variation of a probabilistic Bell measurement for polarised photons, and can be implemented with linear optical elements such as half wave plates and polarising beam splitters. Since the creation of the cluster state occurs before we introduce the quantum computation via measurements, we are at liberty to create the cluster state in a probabilistic manner and purify the result until we have the desired fidelity.

There are two types of fusion gates, type-I and type-II. The first type purports to detect a single photon, leaving the three remaining photons in an entangled linear cluster state. This fusion gate can be described mathematically by the operator
\begin{align}\label{eq:type1}
 \mathcal{F}_{{\rm I},\pm} = \ket{H}\bra{H,H} \pm \ket{V}\bra{V,V}\, ,
\end{align}
where the sign $\pm$ is determined by the polarisation of the photon measured in detector D. Similarly, the type-II fusion gate can be described by the operator
\begin{align}\label{eq:type2}
 \mathcal{F}_{\rm II} = 
 \begin{cases}
 \bra{H,V} + \bra{V,H} & \text{for outcome $(H,H)$ or $(V,V)$ in D1 and D2,}\cr
 \bra{H,H} + \bra{V,V} & \text{for outcome $(H,V)$ or $(V,H)$ in D1 and D2.}
 \end{cases} 
\end{align}
Starting with Bell pairs, the type-II fusion gates clearly cannot grow large clusters on their own since they remove two photons from the entangled state. However, type-II has a much more beneficial behaviour that type-I when the fusion gate fails. The best strategy is therefore to create three-photon entangled states using type-I gates, and subsequently create larger cluster states using only type-II fusion gates. Improvements in the architecture of linear optical quantum computers continue to be made, and in a recent proposal the need for quantum memories is reduced by using a percolation-based \emph{ballistic} approach \cite{gimeno15}.

\section{The future of optical quantum information processing}\label{sec:future}
One of the aims of this review is to show that different quantum information processing tasks have different technological requirements. For quantum computing, we need sources that produce Bell pairs on demand with high efficiency and, perhaps more importantly, identical mode shapes to accommodate the Hong-Ou-Mandel effect. Moreover, the Bell pairs must be very close to pure states. The photodetectors must have a high detection efficiency, so that photon loss in the course of the computation remains low. The linear optical components must similarly be low-loss and accurate. The polarising beam splitters must have nearly perfect transmission or reflection for the polarised photons, and beam splitters must have carefully calibrated transmission coefficients. The exact allowed tolerances of the components of a linear optical quantum computer will be determined by the error correction mechanism that is employed. Finally, the feed-forward nature of linear optical quantum computing means that we require fast, low-loss optical switches. This is currently a major challenge.

An actual implementation of a linear optical quantum computer will not use bulk optical elements, but rather have a chip-based architecture in which microscopic waveguides are wired into programmable circuits. Beam splitters can then be constructed from evanescently coupled waveguides. By adjusting the distance between the waveguides, the transmission coefficient can in principle be carefully calibrated. Recently, photon sources have been placed in or on top of waveguides, which allows for directional coupling of the photon into the waveguide depending on the spin of the photon source \cite{zayats13,lodahl15,coles15}. This new technology can be employed for alternative Bell pair generation methods based on photon which-path erasure and spin readout.

Quantum metrology is similarly challenging to implement. It is known that in the limit of large photon numbers the Heisenberg limit is extremely sensitive to noise \cite{rafal12}. This means that some type of quantum error correction must be employed in order to achieve the Heisenberg limit, and this places the practical challenge on a par with the construction of a full-scale quantum computer. On the other hand, quantum metrology techniques that do not achieve the Heisenberg limit but that nonetheless improve on the shot-noise limit by a constant factor will still be very welcome. Squeezed (quantum) light will be used in the next generation gravitational wave detectors \cite{aligo13}, especially now that gravitational waves have been observed directly \cite{abbott16}.

Quantum communication is arguably the least challenging task to implement in practice. Quantum key distribution requires single-photon sources that may not be fully indistinguishable from each other. However, much care must be taken in the prevention of side-channel detection, in which an eavesdropper can infer or influence the polarisation of a photon via classical methods (e.g., monitoring the photon source for tell-tale signals, etc.). These can be difficult engineering questions that must be solved. Extending quantum communication over longer distances will require quantum repeaters. These devices are much more challenging to build, and require multi-photon entanglement, high-efficiency photodetectors, and generally rather large optical circuits. While repeaters do not have strict fault-tolerance requirements, the techniques that will make them work will likely be similar to those of full-scale quantum computers (indistinguishable photons, fast low-loss switches, etc.).

To conclude, optical quantum information processing presents various physical and engineering challenges for different tasks. Some processes, such as quantum key distribution are currently being implemented in commercial products, while others are still very much in the research stage. I have shown that different tasks have very similar physical requirements at different stages of development, which makes it more likely that as our understanding and mastery of Nature continues, even the more exotic applications will find their way into working devices.

\section*{Acknowledgements}
I would like to thank Nikola Prtljaga for providing me with the data that was used in figure 4.

%\newpage
%\bibliographystyle{tCPH}
%\bibliography{cpr}

\end{document}